\documentclass[final,1p,times]{elsarticle}
\usepackage{lineno,hyperref}
\modulolinenumbers[5]
\usepackage{natbib}
\setcitestyle{authoryear}
\usepackage{amsmath,amssymb,amsfonts}
\usepackage{url}
\usepackage{multirow}
\usepackage[normalem]{ulem}
\usepackage{longtable}
\usepackage{float}
\usepackage{adjustbox}
\usepackage{textcomp}
\usepackage{algorithmic}
\usepackage{textcomp}
\usepackage{csquotes}
\usepackage{tablefootnote}
\usepackage[ruled,vlined]{algorithm2e}
\usepackage{pdfpages}

\journal{Journal of arXiv}

\begin{document}

\begin{frontmatter}
\title{Unleashing the power of disruptive and emerging technologies amid COVID-19}


\author[1]{Sonali Agarwal \corref{mycorrespondingauthor}}
\cortext[mycorrespondingauthor]{Corresponding author}
\ead{sonali@iiita.ac.in}
\author[1]{Narinder Singh Punn}
\ead{pse2017002@iiita.ac.in}
\author[1]{Sanjay Kumar Sonbhadra}
\ead{rsi2017502@iiita.ac.in}
\author[2]{M. Tanveer}
\ead{mtanveer@iiti.ac.in}
\author[1]{P. Nagabhushan}
\ead{p.pnagabhushan@iiita.ac.in }
\author[3]{K K Soundra Pandian}
\ead{soundra.pandian@cca.gov.in}
\address[1]{IIIT Allahabad, Prayagraj, U.P. India-211015}
\address[2]{Discipline of Mathematics, IIT Indore, M.P. India-453552}
\address[3] {Ministry of Electronics \& Information Technology, New Delhi, India}

\author[4]{Praveer Saxena}
\ead{praveersaxena.edu@nic.in}
\address[4]{Ministry of Human Resource Development, Government of India}
\begin{abstract}
 The unprecedented outbreak of the novel coronavirus (COVID-19), during early December 2019 in Wuhan, China, has quickly evolved into a global pandemic, became a matter of grave concern, and placed government agencies worldwide in a precarious position. The scarcity of resources and lack of experiences to endure the COVID-19 pandemic, combined with the fear of future consequences has established the need for adoption of emerging and future technologies to address the upcoming challenges. Since the last five months, the amount of pandemic impact has reached its pinnacle that is altering everyone's life; and humans are now bound to adopt safe ways to survive under the risk of being affected. Technological advances are now accelerating faster than ever before to stay ahead of the consequences and acquire new capabilities to build a safer world. Thus, there is a rising need to unfold the power of emerging, future and disruptive technologies to explore all possible ways to fight against COVID-19. In this  article, we attempt to study all emerging, future, and disruptive technologies that can be utilized to mitigate the impact of COVID-19. Building on background insights, detailed technological specific use cases to fight against COVID-19 have been discussed in terms of their strengths, weaknesses, opportunities, and threats (SWOT). As concluding remarks, we highlight prioritized research areas and upcoming opportunities to blur the lines between the physical, digital, and biological domain-specific challenges and also illuminate collaborative research directions for moving towards a post-COVID-19 world.

\end{abstract}

\end{frontmatter}

\section*{Impact Statement}
Emerging and future technologies recently coined as disruptive technologies, are going to change the world while addressing the present need to combat the COVID-19 pandemic. This article is an effort to facilitate the swift of technological trends among all researchers with its novel applications to create a healthier and sustainable post-COVID-19 world. Specifically, in the present article, we identify the significance of emerging and future technologies and its disruptive appearance worldwide. It is also important to address major challenges involved in possible solutions, the promising directions for long term sustainability, and most importantly, building technology for mankind. Covering the broad range of emerging and future technologies, this  is intended to encourage the policymakers to embrace the convergence of technologies for effective planning, on the other hand, attract the developers to focus on the divergence of technologies to attain application-specific expertise. This study also enlarges the strengths, weaknesses, opportunities and threats of disruptive technologies under the light of COVID-19, so that a safe and prosperous world may get unfolded. 
\section{Introduction}
The rapidly evolving coronavirus disease 2019 (COVID-19) pandemic~\citep{SA1} has changed the life of everyone all over the world with its deadly spread and about 81,278,115 confirmed cases along with 1,774,130 deaths globally as on December 29 2020~\citep{Snew}. The absence of any effective therapeutic agents and the lack of immunity against COVID-19 making the population susceptible. As there are no vaccines available yet, it has now become a global concern and researchers all over the world are trying hard to come over with this pandemic. Due to the present crisis, massive frictional ripples have been created thus various sectors such as tourism, transportation, the supply chain of export and import, healthcare, IT industries, etc., are either reeling under pressure or in standstill phase across the borders. Therefore, there is a constant strain of science and technology to emerge with innovative solutions to mitigate the situation. 

Under COVID-19 crisis, emerging and future technologies are propping up our daily lives to match with the current need and becoming essential day by day. Since human nature is adaptive and necessity is the mother of invention, presently people from all sectors are putting their efforts to explore technological solutions of the problems caused by this global pandemic. Since there are many advances reported in the last decade; apparently, technology-driven applications are establishing critical infrastructure and offering backbone to day-to-day needs of human lives~\citep{vaishya2020emerging}.

While using all common keywords to explore the role of technology to fight against COVID-19 pandemic, it is evident that most of the search results are URLs of bulletins, press releases, white papers, blogs and product catalogues of commercial organisations for awareness and promotion campaign purposes. Although it is interesting to get updated about recent technologies and their advancements; experimental studies and peer-reviewed research papers are essential to build up trust and confidence in these areas. Following this concern, we intend to present the detailed study of prospective techniques and advancements, addressing its suitability to fight against COVID-19. 

\begin{figure}
    \centering
    \includegraphics[scale= 0.3] {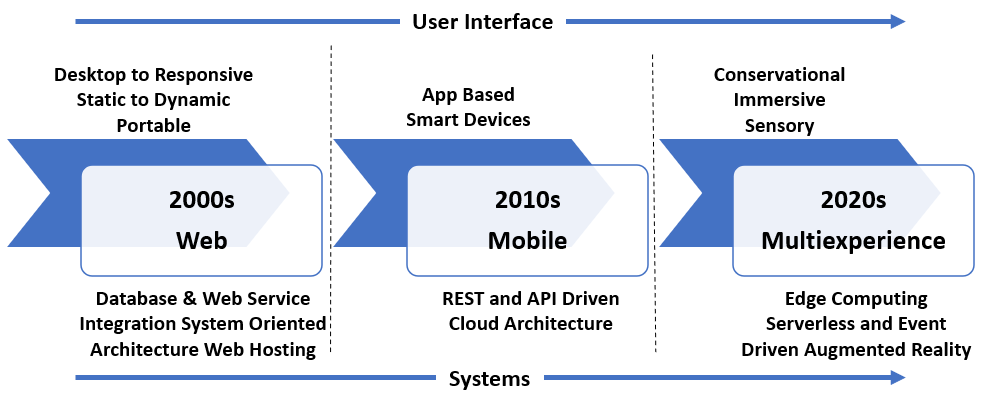}
    \caption{An example of disruptive technology.}
    \label{fig1}
\end{figure}

During the last decade, it is evident that policymakers are taking more interest in the adaptation of novel technologies and this is coming as the operationalization of technological emergence. Emerging technologies are radically novel, relatively growing, and becoming coherent and prominent over time, therefore, have a substantial impact on the socio-economic domain(s)~\citep{rotolo2015emerging}.  As technology adoption rate is speeding up day by day, incremental progress has been witnessed by humans and future technologies are continuously replacing the existing work practices. Disruptive technology is a keyword coined by ~\cite{christensen2000meeting}, which refers to a revolutionary change that uproots an established technology and spawns new industries, markets and value networks that eventually disrupts an existing lifestyle, thereby presenting the power to change the way we work, live, think and behave~\citep{SA5, christensen2015disruptive}. We are using the term \enquote{disruptive technology} as an \enquote{umbrella} that encompasses a novel perception to see the changes in the technological usage pattern with respect to time. We are also trying to map technologies with respect to physical, digital, and biological perspectives in order to support the fight against COVID-19. Fig.\ref{fig1} shows an example of disruptive technology over the past decades.
\begin{figure}
    \centering
    \includegraphics[scale= 0.2] {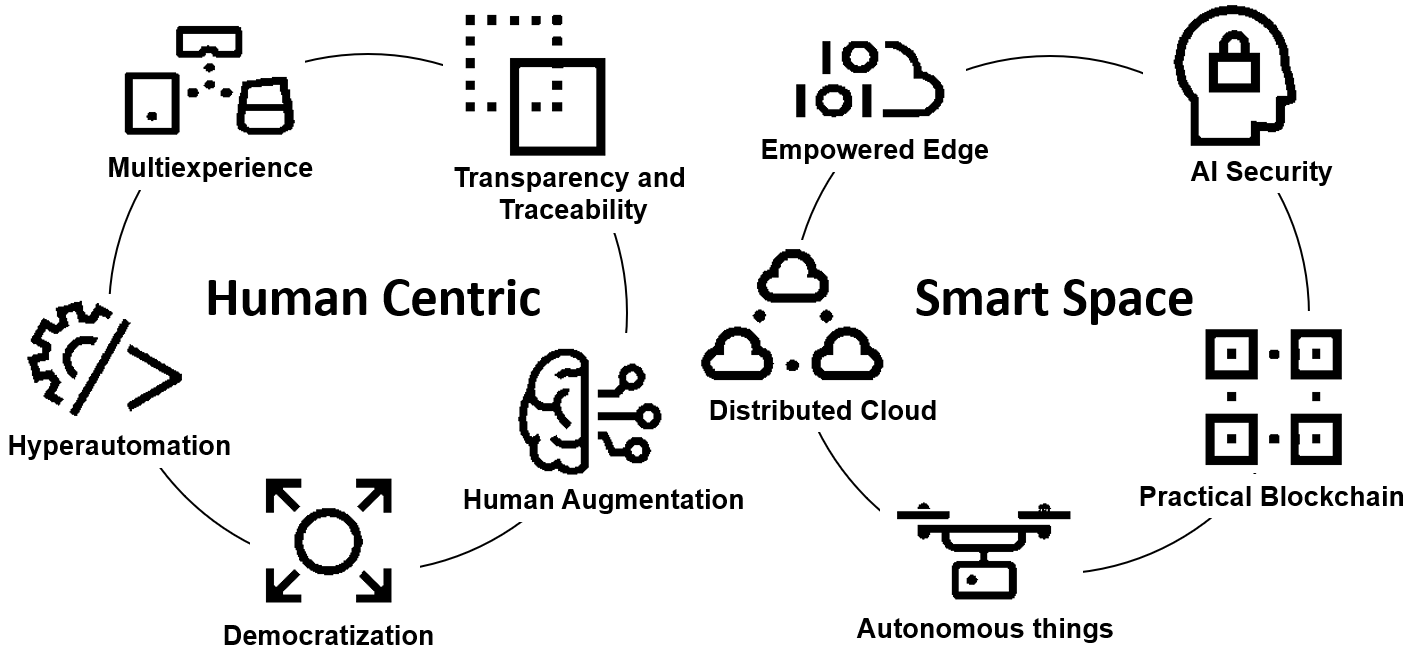}
    \caption{Human-centric and smart-space technologies.}
    \label{fig2}
\end{figure}

As mentioned in the study, if we follow the trails of the disruptive technology landscape, enterprise leaders need to comprehend the nuanced interconnections among people, processes, services, and things. Following from these notions, it is essential to first understand the context~\citep{christensen2015disruptive} before initiating the utilization of any innovative technology. According to Gartner’s recent report~\citep{rotolo2015emerging}, there are certain trends that will drive significant disruption and opportunities over the next 5 to 10 years, especially in the area of infrastructure and operations~\citep{SA7}. A Gartner team member, D. Cearley, highlighted that strategic technology trends now surrounded into two categories: human-centric and smart-space~\citep{fang2020study}. According to an article~\citep{shaw2020understanding}, human-centric technologies can create various opportunities and can drive disruptions in our lifestyle for the future. Human-centric approaches keep people in priority to develop solutions which can strengthen different roles within the organisations such as customers, policymakers, employees, business partners, etc. On the other hand, a smart-space is termed as a physical environment, which offers an open, connected, coordinated and intelligent ecosystem to people for their seamless interaction using the technology-enabled systems along with the more immersive, interactive and automated experience. Fig.~\ref{fig2} highlights the various interconnections between human-centric and smart-space technologies. 

COVID-19 becomes a black swan for the entire world and provokes a very strong wave of change and disruption. The growing impact of the unprecedented human tragedy is not only affecting the people's health but also hampered the functioning of several sectors, including aviation, automotive, pharmaceuticals, banking, consumer electronics, travel and tourism, among others. Thus it is high time to project emerging, future and disruptive technologies to support the above-discussed sectors and also speeding up the battle against the coronavirus. Table~\ref{tab1} represents the different types of disruptive technology along with the supporting emerging technologies.
\begin{table}[]
    \centering
    \caption{Technology categorization}
    \label{tab1}
    \resizebox{\linewidth}{!}{
    \begin{tabular}{|p{1in}|p{1.3in}|p{3.2in}|}
		\hline
		\textbf{Disruptive technology theme} & \textbf{Disruptive technology types} & \textbf{Supporting emerging technologies}\\
		\hline
		\multirow{7}{*}{Human Centric} & Hyperautomation & Machine Learning, Robotics \\ \cline{2-3}
		& Multiexperience & Virtual reality (VR), augmented reality (AR) and mixed reality (MR).\\ \cline{2-3}
		& Democratization & Big Data Analytics. \\ \cline{2-3}
		& Human Augmentation & Virtual Reality, Augmented Reality, Virtual Assistants, Computer Vision, Biometrics. \\ \cline{2-3}
		& Transparency and Traceability & Blockchain and Big Data Analytics, 5G. \\ \hline
		
		\multirow{5}{*}{Smart Space} & Empowered edge & Internet-of-things, Big Data Analytics, 5G \\ \cline{2-3}
		& Distributed Cloud & Cloud Computing, Big Data Analytics. \\ \cline{2-3}
		& Autonomous things & Drones, autonomous vehicles, robotics, 5G. \\ \cline{2-3}
		& Practical Blockchain & Blockchain. \\ \cline{2-3}
		& AI Security & Artificial Intelligence, Blockchain. \\
		\hline
	\end{tabular}}
\end{table}
As an attempt to cover the broad range of emerging, future and disruptive technologies for the purpose to combat COVID-19, following are our contributions:  First, we review the evolving nature of technologies, emphasized on its disruptive nature classified as a human-centric and smart-space theme. Second, we listed down individual technologies and prepared a timeline about their emergence to portray continuous evolutions along with coverage of related research work. Third, we identified technology-enabled COVID-19 use cases in all possible application domains along with thorough coverage of SWOT analysis. Fourth, we summarize the findings which can be adopted easily along with the future recommendations. Lastly, we conclude our findings with some unexplored open research questions to be addressed in the near future by the researchers.

The rest of this paper is systematized as follows: In Section II, we elaborated the background of disruptive technologies and their inclusive areas. In Section III, we presented the origin, conceptual details and research status of various emerging and future technologies. Various use cases based on emerging and future technologies, to combat against COVID-19 are discussed in section IV. SWOT analysis of these technological solutions are also discussed in this section. Next, in section V, we discuss prioritized research areas, common challenges and upcoming opportunities in this context. Lastly, section VI concludes the paper. 

\section{Disruptive technologies} 

The persistent procession of innovative technologies is unfolding in many ways. Almost every development is counted as a breakthrough, and the list of emerging and future technologies are nurturing continuously. While considering the impact of emerging technologies on business or social landscape, it is quite convincing that some technologies actually have the potential to disrupt the status quo, change the people's lifestyle, and reorganize value pools. It is thus necessary that commercial and government organizations should understand the scope of the technologies to cater the upcoming need and prepare accordingly. According to Gartner~\citep{rotolo2015emerging}, technological trends emphasized on the following categories: 
\subsection{Human-Centric technologies }
Human-centric technologies are intended to develop solutions based on user requirements, desires and capacities, keeping prime focus on how individuals can, need and want to perform tasks, rather than expecting users to fine-tune and accommodate their actions to the product. It is one of the most transformative and disruptive technology trends that will relate to how humans interact with the digital world. This method is supported by rigorous testing and engineering practices for the product development that fulfils user requirements and business objectives. Following are the human-centric technological advancements:
\subsubsection{Hyperautomation}
\begin{figure}
    \centering
    \includegraphics[scale= 0.32] {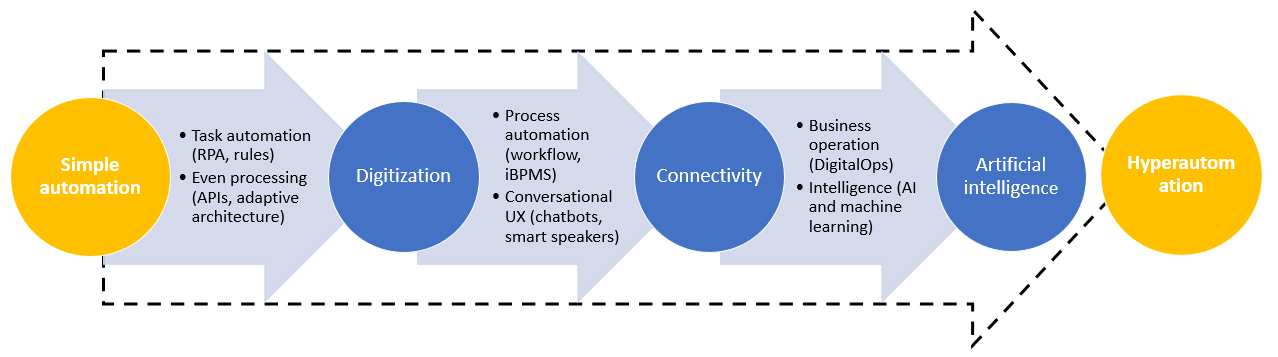}
    \caption{An evolution of hyperautomation.}
    \label{fig3}
\end{figure}

Hyperautomation is a term proposed by Gartner, is one of the human-centric disruptive technologies. It is a combination of multiple machine learning (ML) techniques and artificial intelligence to progressively automate processes and augment humans. It covers all steps of automation such as to discover, analyze, design, automate, measure, monitor and re-assess. It has two components: robotic process automation (RPA)~\citep{van2018robotic} and intelligent business process management suites (iBPMSs)~\citep{dunie2015magic}. Here RPA works as a bridge between legacy and modern systems. It supports knowledge workers on day-to-day routine and eliminates repetitive tasks by means of migrating from the present environment to the next environment with the help of tightly defined integration scripts structure and data manipulation. iBPMSs is suitable for long term processes and works as an unified solution that coordinates people, processes, machines and things. It is a collection of rules to trigger a user interface and manage the contextual work items via robust APIs. It also supports the full life cycle of business processes such as discovery, analysis, design, implementation, execution, monitoring and continuous optimization, to facilitate collaboration among citizens and professionals on iterative development and improvement of processes and decision models. Fig.~\ref{fig3} shows the evolution of hyperautomation.

\subsubsection{Multiexperience}
\begin{figure}
    \centering
    \includegraphics[scale= 0.35] {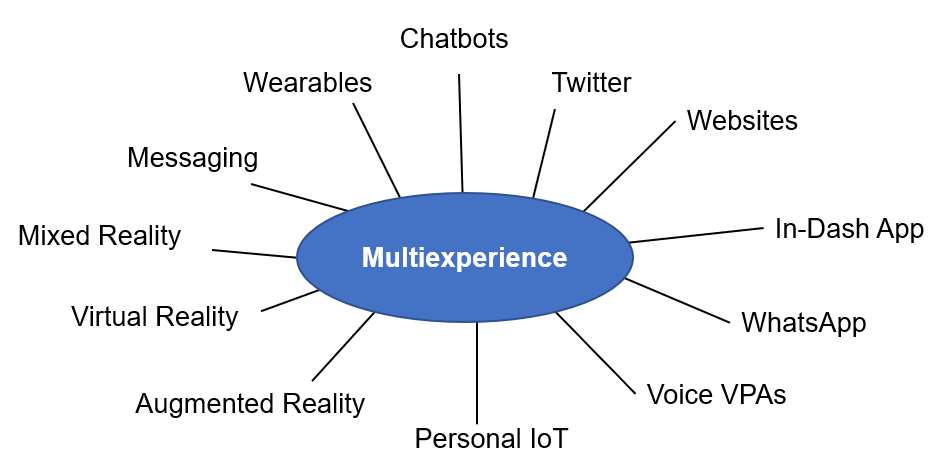}
    \caption{Multisensory support devices for multiexperience.}
    \label{fig4}
\end{figure}

According to Gartner, in the next decade, the users will experience a noteworthy change in their exposure to the digital world and its interaction techniques. It offers moving from the conversational platforms composed of technology-literate people to people-literate technology such as virtual reality (VR), augmented reality (AR) and mixed reality (MR) in a revolutionary way to perceive the digital world~\citep{milgram1994taxonomy, farshid2018go}. In this way, it offers a cohesive change in both perception and interaction models which leads to the future, based on multimodal experience and empowered by multisensory supports to provide a richer environment for delivering nuanced information as shown in Fig.~\ref{fig4}.
\subsubsection{Democratization}
\begin{figure}
    \centering
    \includegraphics[scale= 0.35] {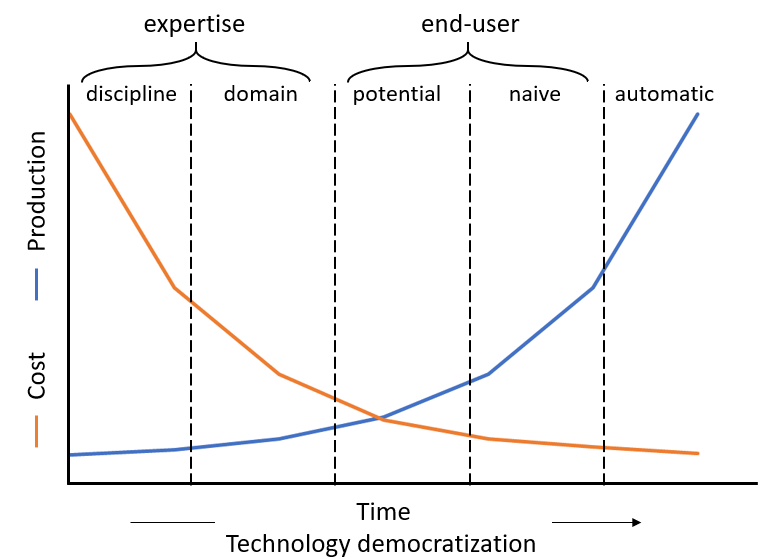}
    \caption{Democratization of technology over the time of its emergence.}
    \label{fig5}
\end{figure}

Democratization is an intensive way to offer people an easy access to technical and business expertise by means of radically simplified \enquote{citizen access} experience and with minimum training~\citep{garg2020democratization}. As mentioned in the report, Gartner emphasizes democratization on four levels such as data, analysis, design, development and knowledge. In data and analysis democratization tools are designed to target the development community; in development democratization AI tools are utilized for application development; in design democratization automated applications are developed from the user’s perspective; and in knowledge democratization application dimensions are further expanded. Fig.~\ref{fig5} describes the democratization of technology starting from its emergence representing that the cost of technological development drops whereas its production rate increases with the passage of time, which is divided into phases as expertise in discipline (core technology) and domain (specific domain), potential (with technical skills) and naive end-users, and finally developed technology operates automatically with the users' intervention.
\subsubsection{Human augmentation}
\begin{figure}
    \centering
    \includegraphics[scale= 0.30] {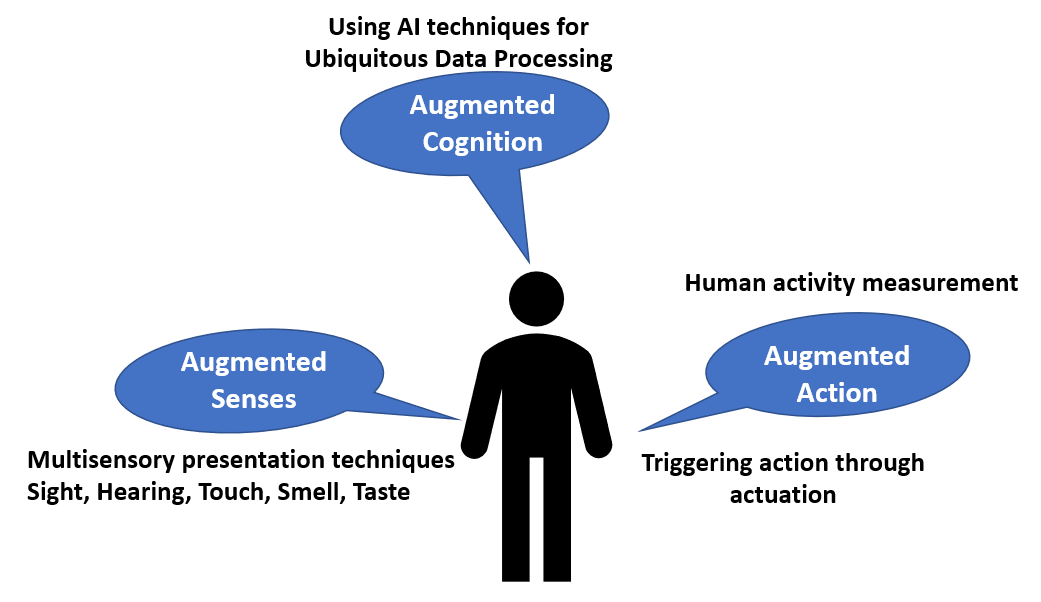}
    \caption{Components of physical augmentation.}
    \label{fig6}
\end{figure}

Human augmentation is an area that discovers the way of enhancing cognitive and physical components as an integral part of human experience (shown in Fig.~\ref{fig6}). This is indicating the consumerization effect~\citep{raisamo2019human}, where workforces can get enhanced capabilities to contribute to a better working environment. In physical augmentation~\citep{unknown}, people host technological components on their bodies to improve human capabilities such as a wearable device. These components are classified as sensory augmentation (hearing, vision, perception) and biological function augmentation (exoskeletons, prosthetics). In cognitive augmentation, human abilities can be utilized for better thinking, decision making,  learning and experiencing new things. Cognitive augmentation~\citep{10.3389/fnhum.2019.00013} correspondingly comprises brain augmentation which can be physically implanted that deals with intellectual reasoning, and genetic augmentation which covers somatic gene and cell therapy.

\subsubsection{Transparency and traceability}
The evolving technologies are raising trust issues and due to this users are now becoming more concerned about their privacy and seeking secure and controlled sharing of data~\citep{veloso20202020}.  Many organizations are realizing that managing personal data of the users is a sensitive issue and needs strict policies about its fair use. To address this genuine requirement, transparency and traceability are most essential to support these digital ethics and privacy needs (Fig.~\ref{fig6}). It refers to a set of initiatives, actions and supporting technologies along with the practices intended to address regulatory requirements, preserve an ethical approach to use artificial intelligence (AI) and supplementary progressive technologies, and restore the rising absence of trust in companies. To build out transparency and trust practices, following three areas are in prime focus~\citep{8936158}: (1) AI and ML; (2) personal data privacy, ownership and control; and (3) ethically aligned design. As consumers become more aware of how their data is being collected and used, organizations are also recognizing the increasing liability of storing and gathering the data.

\begin{figure}
    \centering
    \includegraphics[scale= 0.25] {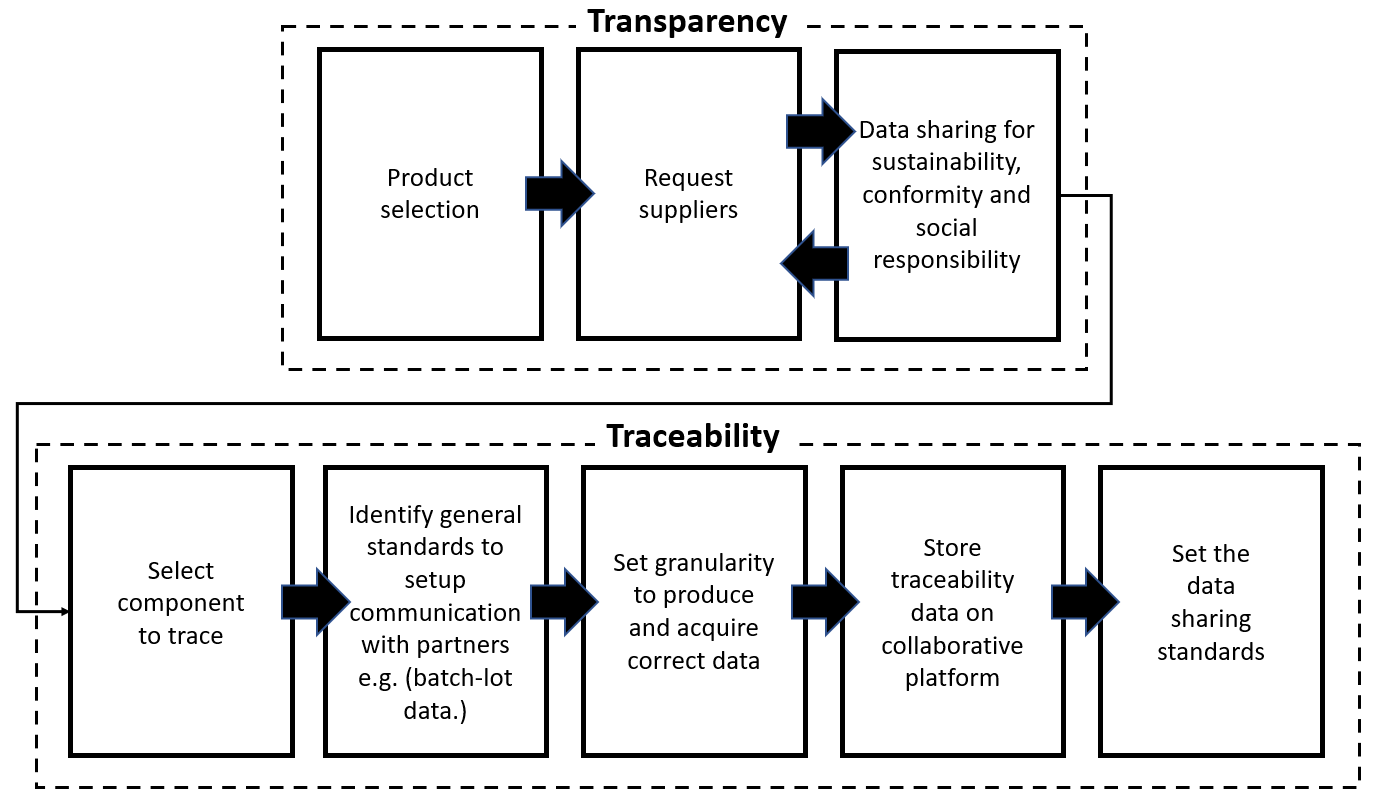}
    \caption{Chain of transparency and traceability.}
    \label{fig7}
\end{figure}

\subsection{Smart-space based technologies}
\begin{figure}
    \centering
    \includegraphics[scale= 0.45] {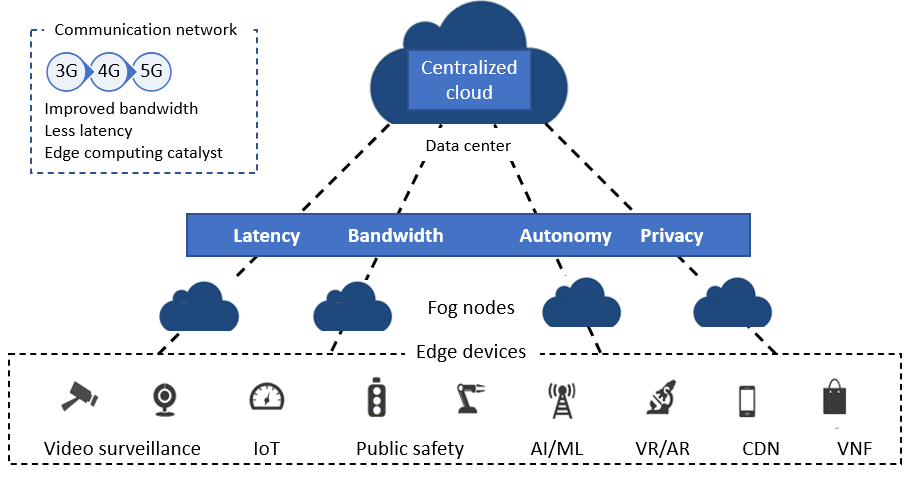}
    \caption{Components of smart-space based technology.}
    \label{fig8}
\end{figure}

Earlier, ad hoc and wireless communication technologies were used to develop an information-rich working environment which is now replaced by smart-space, that is based on ubiquitous computing of common objects to facilitate context-aware services to the users in smart living environments~\citep{el2020energy}. It covers design, implementation and evaluation of information domains supported by novel architectures to deliver intelligent services for living society. Fig.~\ref{fig8} shows typical components of smart-space technology.
\subsubsection{Empowered edge}
Edge computing is a revolutionary area of research in which data acquisition and processing are placed closer to the sources, repositories and end-users to keep the low transmission overhead and reduced latency~\citep{ji2020artificial}. This approach is closely related to IoT systems and is extended to related techniques such as robots, drones, autonomous vehicles and operational systems, and supports device democracy. According to Gartner, by the end of 2023, approximately 50\% of the companies will migrate from cloud computing and adopt the edge computing models.

\subsubsection{Distributed cloud}

A distributed cloud is an advancement over cloud computing in which it utilizes geographically dispersed infrastructure that mainly runs services at the edge level. Instead of storing the data at a centralized repository, it works on the distribution of services at different locations with well-defined policies for operation, governance, updates and evolution of the services. It is fully equipped with computation, storage, and networking capabilities in the form of micro-cloud which is located outside the centralized system. Fog computing~\citep{habibi2020fog} and edge computing~\citep{cao2020overview} are different categories of the distributed cloud. Establishing a distributed cloud comprises data processing computing closer to the end-user, decreased latency and enhanced security in a real-time environment. A distributed cloud also processes data in real-time.

\subsubsection{Autonomous things}

Autonomous things are interconnections of AI-enabled physical devices that are designed to replace humans for many operations such as robots, drones, autonomous vehicles/ships and appliances. It is more than automation because it exploits AI to deliver advanced services to the people~\citep{langley2020internet}. As technology competence is growing day by day, regulatory measures and social acceptance is increasing, deployment of autonomous things will be trending in various societal applications. As autonomous things are flourishing, there is a clear shift from stand-alone intelligent systems to a swarm of collaborative intelligent systems with multiple workable devices.

\subsubsection{Practical blockchain}

Blockchain consists of a secure distributed ledger, a chronologically ordered list of encrypted signatures, unalterable transactional records shared by all participants in a network. Blockchain is secure because it allows parties to trace assets back to their origin in any supply chain and allows safe interaction between two or more parties in a digital environment and easy exchange of money without any central authority. For example, in a smart contract payment, events can automatically get triggered after receiving the goods. In many ways, blockchain is now reshaping the industries by empowering trust, transparency and value exchange across the business ecosystems and potentially lowering costs and latencies~\citep{gatteschi2020blockchain}. There are five integral parts of a blockchain model: A shared and distributed ledger, immutable and traceable ledger, encryption, tokenization and a distributed public consensus mechanism~\citep{banerjee2020study}. However, blockchain is not feasible to implement with many business operations due to many technical deficiencies such as poor scalability and interoperability~\citep{fabiano2017internet}.

\subsubsection{AI security}

\begin{figure}
    \centering
    \includegraphics[scale= 0.45] {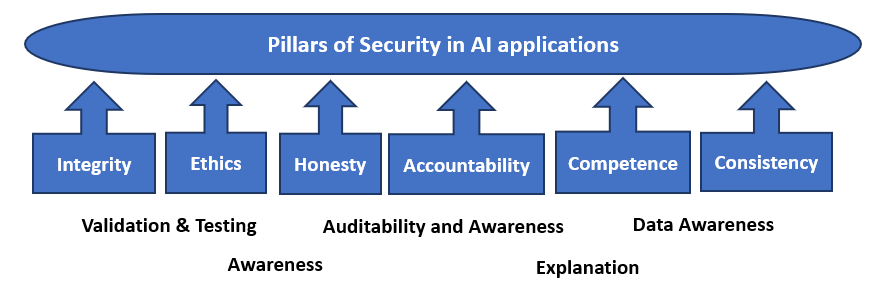}
    \caption{AI security systems.}
    \label{fig9}
\end{figure}

AI and ML are rapidly integrated into human decision making by means of several use cases such as in-home voice assistants. self-driving cars, business analytics and many more. It is playing a crucial role in many technological advancements and becoming challenging day by day due to possibilities of attack in IoT, cloud computing, microservices and highly connected systems in the smart-space. Therefore it is urgently needed that security and risk should be of prime focus especially with AI-enabled services on three key areas: (1) safety and protection of AI-powered systems, (2) empowering AI to enhance security defence, and (3) estimating nefarious use of AI by attackers. Fig.~\ref{fig9} shows the details of AI security systems.

\section{Emerging and future technologies}
Emerging technologies highlight the technical innovations in various fields while imposing positive impacts which can benefit the living society. On the other hand, futures technology is the systematic study of technical advancements and innovations that tends to highlight the purpose of predicting how people will reside in the future. Due to rapid technological advancement in innovations, it is always difficult to realise the fine gap between these two phases (emerging and future) of technology evolution. After extensive behavioural analysis, all technologies discussed in the present article are classified in following three categories (as shown in Fig.~\ref{fig30}): 
\begin{figure}
    \centering
    \includegraphics[scale= 0.25] {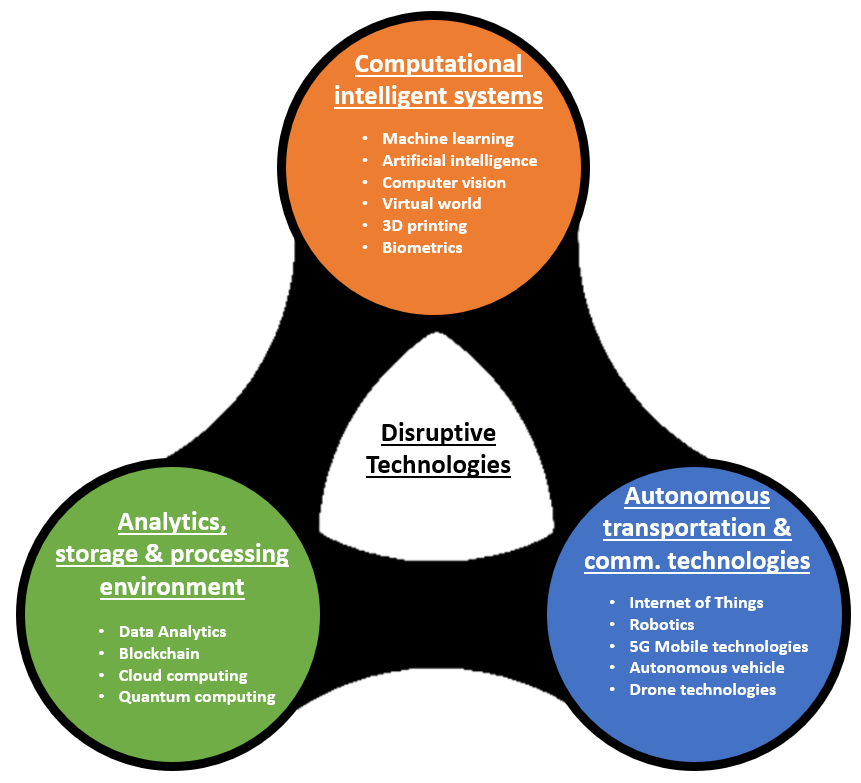}
    \caption{Categories of disruptive technologies.}
    \label{fig30}
\end{figure}
\begin{itemize}
    \item {} Computational intelligent systems.
    \item {} Analytics, storage and processing environment.
    \item{} Autonomous transportation and communication technologies.
\end{itemize}
In the subsequent subsections, the background details of all these emerging and future technologies are covered. 
\subsection{Computational intelligent systems}
\subsubsection{Machine learning}
Machine learning approaches  transformed the world of data analysis in the past few decades. The learning is possible in supervised, semi-supervised and unsupervised modes. Massive data about COVID-19 is generated incrementally once it was declared as pandemic by WHO. Machine learning algorithms are playing a vital role in COVID-19 diagnosis,  infection prediction, recovery rate, death ratio, surveillance, etc. Machine learning based recommender systems could play a robust and efficient method to control the spread of coronavirus. 

\begin{figure}
    \centering
    \includegraphics[scale= 0.3] {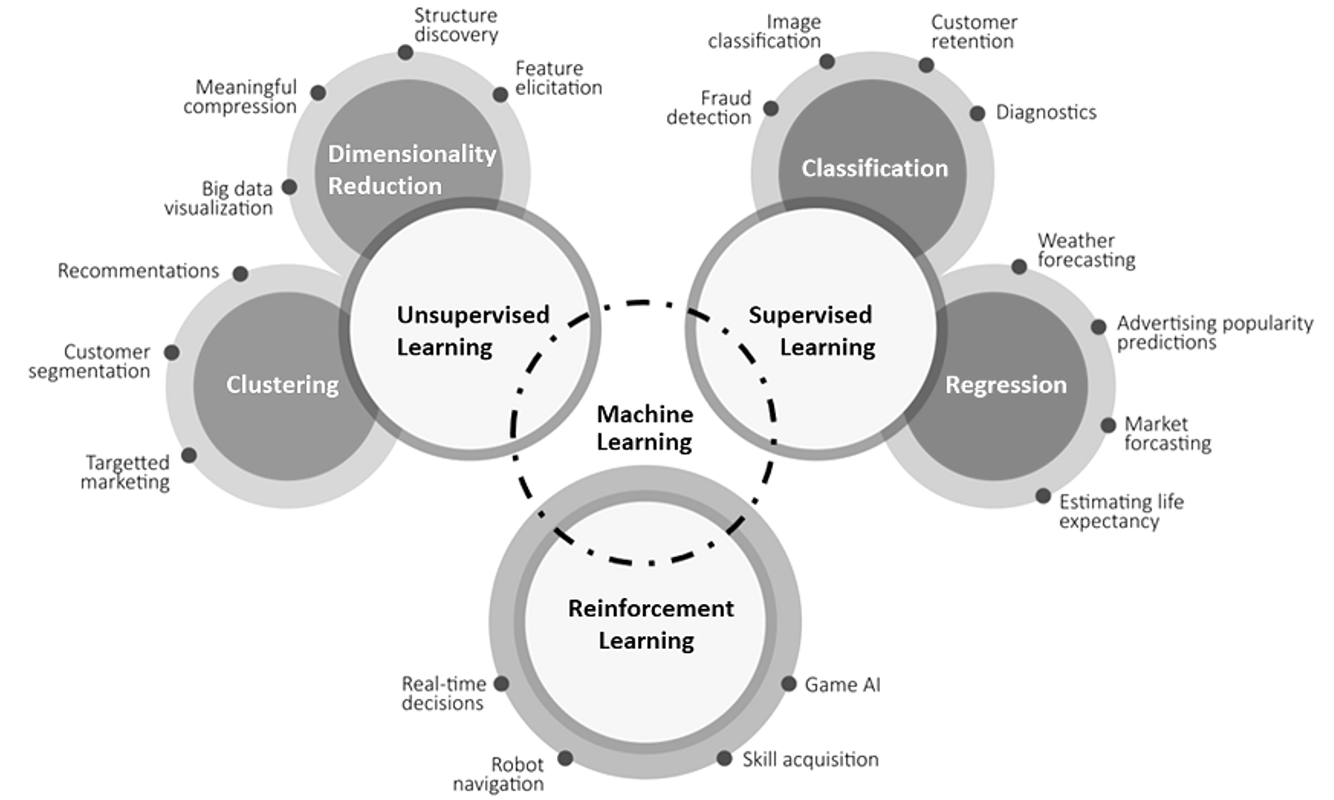}
    \caption{Branches of machine learning and application domains.}
    \label{fig13}
\end{figure}

\begin{figure}
    \centering
    \includegraphics[scale= 0.35] {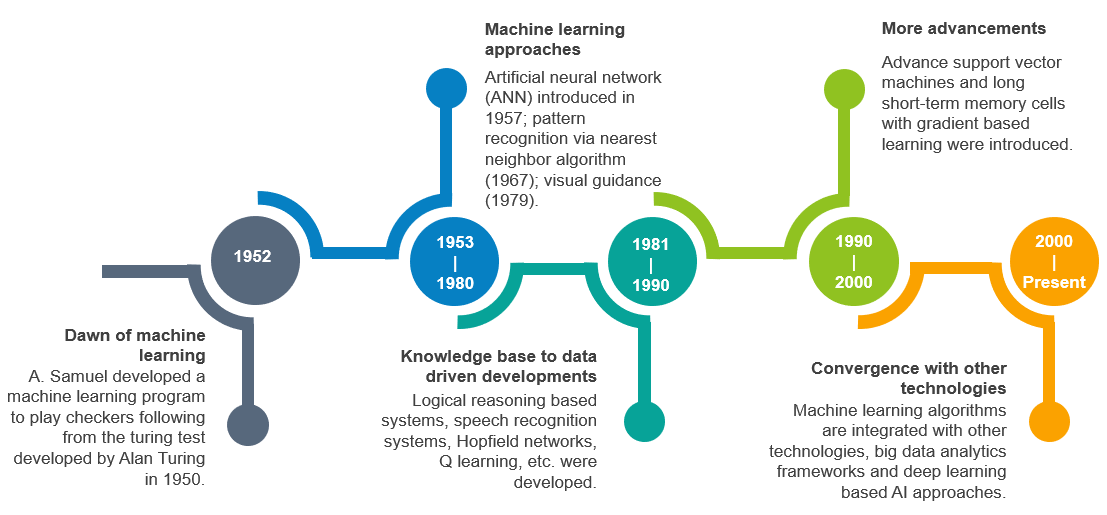}
    \caption{Evolution of machine learning algorithms.}
    \label{fig14}
\end{figure}

Since the origin of machine learning in 1959 by~\cite{samuel1959some}, the remarkable work in this field has revolutionized the world we see today. It is the broad area of study in the field of information technology that aims to extend the program’s ability to learn from its past experience without explicitly coding again. Since there is nothing like an alpha algorithm that can solve everything, Fig.~\ref{fig13} presents the categories of machine learning schemes developed so far along with their respective application domains. Each of these categories tend to deliver solutions for real-world problems. Fig.~\ref{fig14} describes the evolution of machine learning approaches over the years. In Table~\ref{tab5} recent surveys are presented describing the state-of-the-art works proposed for application specific domains.

\subsubsection{Artificial intelligence (AI)}
\begin{figure} [h!]
    \centering
    \includegraphics[scale= 0.35] {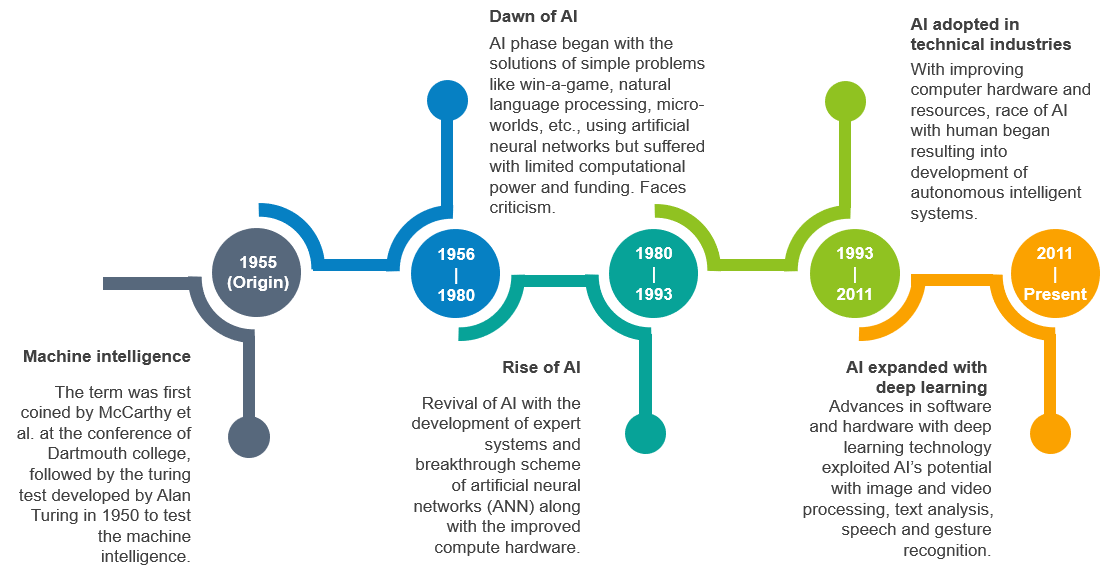}
    \caption{Evolution of AI since its origin.}
    \label{fig10}
\end{figure}

As the COVID-19 eruption continues to spread across the globe, researchers are adopting artificial intelligence as a way to address the challenges of the pandemic. Ranging from virtualized manufacturing of advanced drugs to apply data analytics for rapid contact tracing, all such disruptions can be significantly mitigated through AI. It is also true that AI is not going to replace human experts completely but it has certainly turned out to be a useful tool to monitor and respond to the crisis with the help of its powerful algorithms and tools. 
Artificial intelligence also described as machine intelligence was coined by~\cite{mccarthy2006proposal} in 1955, is an emerging area of study that aims to simulate the real-world problems into the binary world (Information technology), with the aim to maximize the likelihood of decisions to achieve the desired goals that mimics the human brain. Fig.~\ref{fig10} highlights the disruptive findings, illustrating the strength of AI since its origin. AI covers a wide range of research approaches to aid in the development of machine intelligent systems e.g. cognitive science, artificial neural networks, faster computing hardware, algorithmic complexity theory, robotics, swarm intelligence, embodied and logic-based systems which has been utilized by more than 77\% of the consumers~\citep{NP201}. Table~\ref{tab2} highlights literature surveys and recent research progress in the field of AI. Over the last couple of decades, AI concepts have been evolving and are achieving better recognition from academics, government, industry and other areas.
 \begin{table}[h!]
    \centering
    \caption{Recent surveys conducted in the field of machine learning.}
    \label{tab5}
     \resizebox{\linewidth}{!}{
    \begin{tabular}{|p{1.2in}|p{1.4in}|p{0.25in}|p{2.0in}|}
        \hline
         \textbf{Authors}                          & \textbf{Title}                                                           & \textbf{Year} &  \textbf{Main contribution}                                                                                                                       \\
         \hline
\cite{zhang2020machine}    & Machine learning testing: Survey, landscapes and horizons & 2020 & A comprehensive survey on machine learning testing research while addressing the correctness, robustness and fairness of the results. \\ \hline
\cite{wang2019machine}        & Machine learning for survival analysis: A survey                & 2019 & Presents the comprehensive survey of the survival analysis with censored data and its challenges using machine learning algorithms.       \\ \hline
\cite{mehrabi2019survey}    & A survey on bias and fairness in machine learning               & 2019 & Analyses the discriminatory behavior of the models in the sensitive decision-making environment.                                          \\ \hline
\cite{mahdavinejad2018machine} & Machine learning for Internet of Things data analysis: A survey & 2018 & Considering smart cities as a major concern, the author addresses the challenges associated with the IoT via machine learning approaches. \\ \hline
\cite{allamanis2018survey}   & A survey of machine learning for big code and naturalness       & 2018 & Highlights the intersection of software engineering and machine learning to exploit its programming patterns.                             \\ \hline
\cite{qiu2016survey}          & A survey of machine learning for big data processing            & 2016 & Presents the machine learning perspective of big data with discussion on challenges and its solutions.                                    \\ \hline
    \end{tabular}}
\end{table}
\begin{table}[h!]
    \centering
    \caption{Survey papers on AI to illustrate its research progress in various domains.}
    \label{tab2}
    \resizebox{\linewidth}{!}{
    \begin{tabular}{|p{1.2in}|p{1.4in}|p{0.25in}|p{2.0in}|}
        \hline
         \textbf{Author}                     & \textbf{Title}                                                                          & \textbf{Year} & \textbf{Main contribution}                                                                                               \\
         \hline
\cite{shi2020artificial}     & Artificial Intelligence for Social Good: A Survey                              & 2020 & Proposes ethical factors and AI4SG case studies for future initiatives.                                         \\  \hline
\cite{tong2019artificial}    & Artificial Intelligence for Vehicle-to-Everything: A Survey                    & 2019 & Implication of AI driven approaches to aid in the internet of vehicles to everything issues in various domains. \\  \hline
\cite{bannerjee2018artificial} & Artificial Intelligence in Agriculture: A Literature Survey                    & 2018 & Category based AI advancements to encounter agriculture challenges.                                             \\  \hline
\cite{li2017applications}      & Applications of artificial intelligence in intelligent manufacturing: a review & 2017 & Proposes intelligent manufacturing framework based on AI advancements.                                          \\  \hline
\cite{muller2016future}   & Future progress in artificial intelligence: A survey of expert opinion         & 2016 & Raises humanity risks concerns for the upcoming AI driven super intelligent systems.\\  \hline
   
    \end{tabular}}
\end{table}

\subsubsection{Computer vision}
The COVID-19 contamination has triggered an urgent need to contribute in the combat against an enormous menace to the human population. As the virus is extremely easy to spread, the top safety measure is to limit interaction with others to follow social distancing. Computer vision, as a subfield of artificial intelligence, has gained attention among researchers as a strong tool to solve complex problems like monitoring, tracking, medical image analysis, etc. to aid fight against COVID-19 spread.

\begin{figure}[H]
    \centering
    \includegraphics[scale= 0.4] {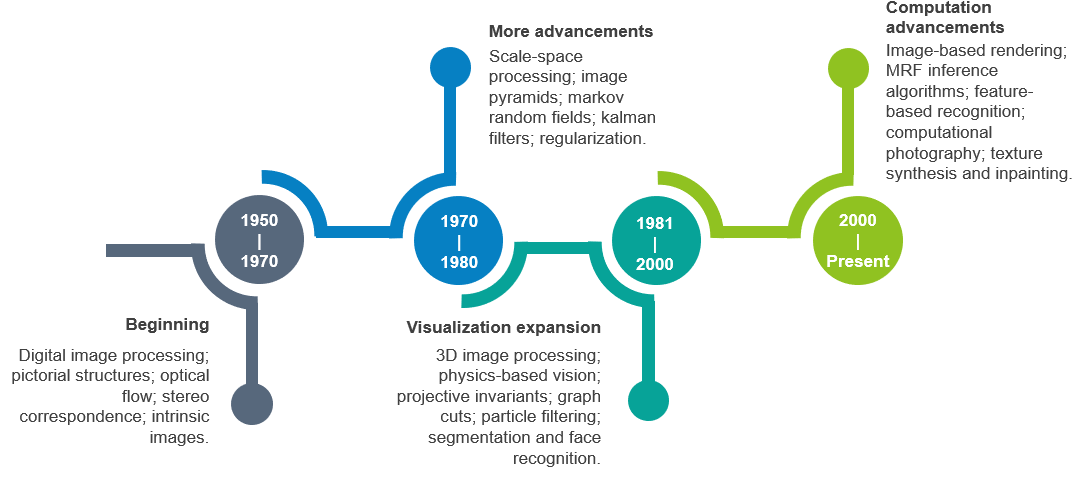}
    \caption{Disruptive technologies in computer vision since its origin.}
    \label{fig15}
\end{figure}

\begin{table}[H]
    \centering
    \caption{Recent surveys conducted in the field of computer vision.}
    \label{tab6}
    \resizebox{\linewidth}{!}{
    \begin{tabular}{|p{1.2in}|p{1.4in}|p{0.25in}|p{2.0in}|}
    \hline
         \textbf{Author}                     & \textbf{Title}                                                                                     & \textbf{Year} & \textbf{Main contribution}                                                                                                                                            \\ \hline
\cite{ulhaq2020computer}  & Computer Vision for COVID-19 Control: A Survey         & 2020 & Explores the recent contributions in COVID-19 research with the help of computer vision approaches. \\ \hline  
\cite{zou2019object2}   & Object detection in 20 years: A survey   & 2019 & Presents the technical overview and comparative analysis of the objection approaches.                                                                        \\ \hline
\cite{tan2018survey}     & A survey on deep transfer learning                                                        & 2018 & Explores the existing deep learning approaches that can be fine-tuned to achieve the desired results.                                                        \\ \hline
\cite{zhang2018deep}   & Deep learning for sentiment analysis: A survey                                            & 2018 & Illustrates the recent research works for semantic analysis via deep learning approaches.                                                                    \\ \hline
\cite{hohman2018visual}  & Visual analytics in deep learning: An interrogative survey for the next frontiers         & 2018 & Highlights the role of visual analytics in the study of deep learning using the human centric interrogative framework that involves the use of 5 Ws and how. \\ \hline
\cite{salahat2017recent} & Recent advances in features extraction and description algorithms: A comprehensive survey & 2017 & Addresses the recent advancements for feature extraction and description algorithms in the area of computer vision.                                          \\ \hline
    \end{tabular}}
\end{table}

Computer vision, emerged from artificial intelligence, is the study of analysis of images and videos to gain deep understanding to serve the purpose of a concerned task. Starting from the distinguishing handwritten text from typed text in the 1950s, it has grown exponentially. This field studies the image classification, object detection, object localicalization, object segmentation, image captioning, etc. This field is also treated as if it provides eyes to the computers~\citep{ibrahim2020understanding}. Fig.~\ref{fig15} describes the disruptive technologies developed in this field of study since its origin. It is composed of several image and video processing tasks with a wide area of applications in industrial inspection or quality control, surveillance and security, face recognition, gesture recognition, medical image analysis, virtual reality, and a lot more~\citep{wiley2018computer}. With such wide real-world applications of this field, most research communities are exploring its dimensions to understand its potential. Table~\ref{tab6} presents the recent survey articles that can be utilized to explore its application domain.

\subsubsection{Virtual world}
As the lockdown period is going long and curfews are imposed for precautionary measures in many areas, it is bringing localities to a standstill. Augmented reality (AR) and virtual reality (VR) applications are emerging as a powerful tool to accomplish remote socialization, illustrating experience of virtual shopping, events and conducting business meets. It is not only offering day-to-day, people-to-people interactions with minimum contact but also satisfies the consumers’ expectations of real world experience without any outer exposure and investment in terms of time and money. With this application, it is evident that this technology could aid in following the set government protocols while also continuing the personal and daily business requirements. 
\begin{figure}[H]
    \centering
    \includegraphics[scale= 0.4] {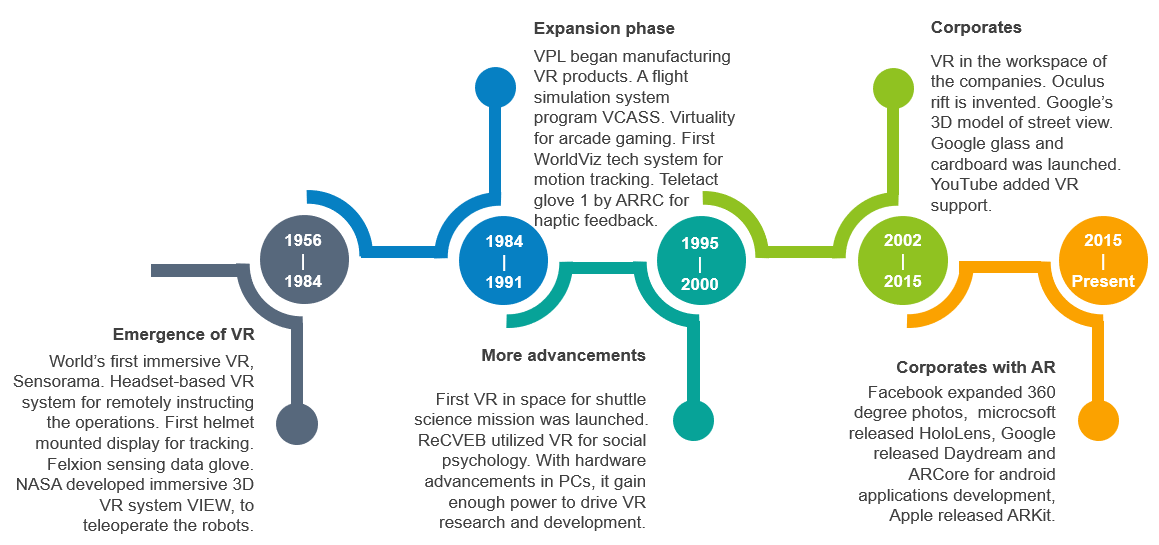}
    \caption{Timeline of virtual reality.}
    \label{fig22}
\end{figure}
Virtual reality (VR) technology was invented in 1957 by Morton Heilig as a multimedia device called the Sensorama~\citep{heilig1962sensorama}. However, the term ‘virtual reality’ was coined much later in 1987 by researcher Jaron Lanier. Virtual reality (VR)~\citep{alturki2019augmented} is a mechanism to create a simulated environment using headsets that blend out the real world, instead of watching on a display, immersing a person in a digital 3D environment. Unlike conventional user display interfaces, VR places users inside an experience. By simulating several senses such as audio, video, touch, etc., the computer is transformed into a gatekeeper to this artificial world. Unlike VR, augmented reality (AR) overlays digital objects onto the real world via smartphone screens or displays, it does not create the whole artificial environment to replace real with a virtual one whereas the mixed reality (MR) is an extension of AR, where users can interact with digital objects placed in the real world (e.g. playing a holographic piano located in the room via an AR headset). Augmented means to add or enhance something and in the case of augmented reality, graphics, sounds, and touch feedback are added in real experience. Virtual intelligence (VI)~\citep{luck2000applying} is an advanced term given to artificial intelligence that applies to the virtual world. It is a sophisticated program designed to provide convenience to modern computer users. Unlike conventional AI, VIs are only utilized to assist the user and process data. Some VIs may have ‘personality imprints’, with their behavior parameters, speech pattern and appearance based on specific individuals, although it is illegal to make VIs based on currently living people. Extended reality (XR) is a term referring to all real-and-virtual combined environments and human-machine interactions generated by computer technology and wearables, where the \enquote{X} represents a variable for any current or future spatial computing technologies~\citep{chuah2018and}. XR most commonly includes virtual, augmented, and mixed reality.  Fig.~\ref{fig22} shows the incremental growth in VR technology whereas the literature papers are considered in Table~\ref{tab12}. The development of VR/AR/XR must be incorporated with appropriate legal compliances.

\begin{table}[]
    \centering
    \caption{Survey papers on advancements in virtual world techniques.}
    \label{tab12}
    \resizebox{\linewidth}{!}{
    \begin{tabular}{|p{1.2in}|p{1.4in}|p{0.25in}|p{2.0in}|}
   \hline
         \textbf{Author}                       & \textbf{Title}                                                                                                                             & \textbf{Year} & \textbf{Contribution}                                                                                       \\ \hline
\cite{eiris2020desktop}    & Desktop-based safety training using 360-degree panorama and static virtual reality techniques: A comparative experimental study  & 2020 & Explores the relationship between realism and hazard identification in 360-degree panorama and static virtual reality however 360-degree panorama  present difficulties for hazard identification as compared to VR.                                                                 \\ \hline       
\cite{el2019survey}     & Survey on depth perception in head-mounted displays: distance estimation in virtual reality, augmented reality, and mixed reality & 2019 & Contributions made to improve depth perception and specifically distance perception are discussed. \\ \hline
\cite{fan2019survey}       & A survey on 360 video streaming: Acquisition, transmission, and display                                                           & 2019 & Discussed the 360-degree head-mounted VR system.                                                   \\ \hline
\cite{anbarjafari2019virtual} & Virtual Reality and Its Applications in Education: Survey                                                                         & 2019 & Discussed the usability of 3D reality in education.                                                \\ \hline
\cite{giap2019virtual}    & Virtual reality medical application system   & 2019 & Applied VR for patient monitoring.                                                                 \\ \hline
    \end{tabular}}
\end{table}

\subsubsection{3D printing}
Presently healthcare systems are at high risk and due to increasing demand, system capacity has to be elevated in order to maintain the supply of medical hardwares (face masks, ventilators, and breathing filters, PPE kits, bubble helmets, etc.). Thus, it is essential that all around the world, the government should take extreme measures to boost productivity and enhance the supply of essential medical equipment. To fulfil these demands, three-dimensional (3D) printing can contribute positively as a disruptive digital manufacturing technology for better production and subsequently encourage the effort of hospital workers in the middle of this emergency.
\begin{figure}
    \centering
    \includegraphics[scale= 0.4] {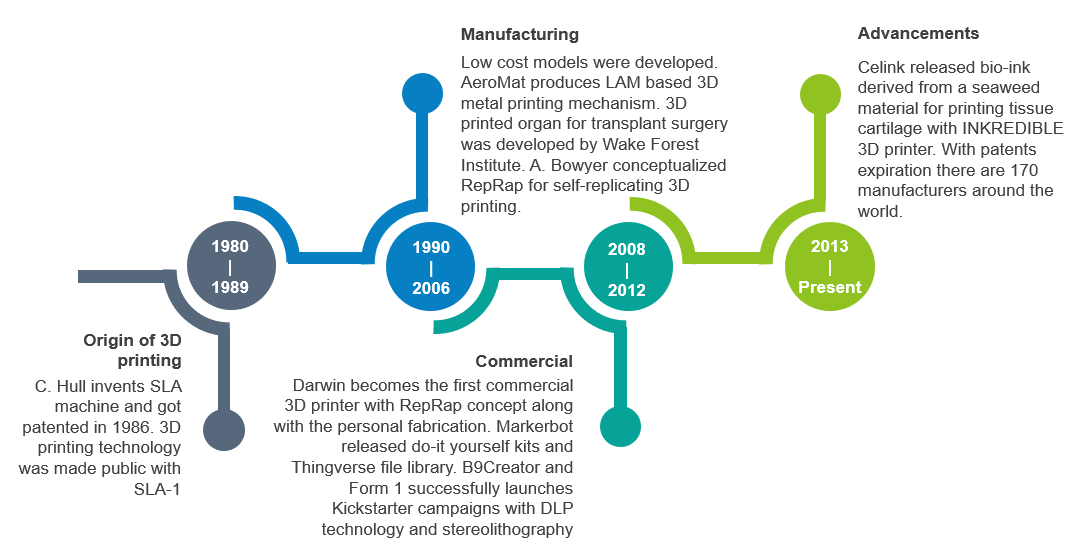}
    \caption{Timeline of 3D printing Technology.}
    \label{fig23}
\end{figure}
\begin{table}[H]
    \centering
    \caption{Survey papers on 3D printing.}
    \label{tab13}
    \resizebox{\linewidth}{!}{
    \begin{tabular}{|p{1.2in}|p{1.4in}|p{0.25in}|p{2.0in}|}
    \hline
         \textbf{Author}                   & \textbf{Title}                                                                      & \textbf{Year} & \textbf{Contribution}                                                                                                               \\ \hline
\cite{huang2020survey}      & A survey of design methods for material extrusion polymer 3D printing      & 2020 & Explores the design methods for material extrusion polymer 3D printing along with the state-of-the-art approaches and research gaps to develop future possibilities.  \\  \hline         
\cite{olsson20193d} & 3D-Printing Technology in Construction: Results from a Survey              & 2019 & Investigated the implementation of 3D-printing in the construction sector.                                                  \\ \hline
\cite{aimar2019role} & The role of 3D printing in medical applications: a state of the art        & 2019 & 3D-printing Discussed the usefulness, drawbacks and how powerful technology the 3D printing is in the medical domain.      \\ \hline
\cite{durfee2019medical} & Medical applications of 3D printing                                        & 2019 & How 3D printing has been (and could be) used in various medical applications.                                              \\ \hline
\cite{matias20153d} & 3D printing: On its historical evolution and the implications for business & 2015 & Describes the historical evolution of additive manufacturing technologies and influencing factors involved in 3D printing. \\ \hline
    \end{tabular}}
\end{table}

In 1974, David E. H. Jones laid out the concept of 3D printing~\citep{horvath2014brief}. It is a technique in which computers are capable of creating three-dimensional objects with raw materials (such as liquid molecules or powder grains being fused together)~\citep{ntousia20193d}. Originally, it is a process that deposits a binder material onto a powder bed with inkjet printer heads layer by layer. Presently, the use of 3D printing is limited and majorly used in both rapid prototyping and manufacturing. The desired objects can be of any geometry or shape and are typically produced using digital model data from a 3D model or another electronic data source such as an additive manufacturing file (AMF) (usually in sequential layers)~\citep{roebuck20113d}. This process is the opposite of subtractive manufacturing which is cutting out / hollowing out an object through machinery, which is capable of producing complex objects using less material compared to conventional manufacturing methods~\citep{gibson2014additive}. 3D printing is useful in every application domain such as medical infrastructure, drug discovery, defence system, industrial production, etc. Fig.~\ref{fig23} highlights the timeline of 3D printing technology and Table~\ref{tab13} considers related survey papers for discussion.

\subsubsection{Biometrics}
For monitoring, infrared thermometers along with facial and iris recognition systems are increasingly being used as most desirable contactless techniques during this pandemic.  Due to the infectious COVID-19 disease, globally, public and private organizations have stopped the use of contact-based biometric systems. Contactless biometric systems are not only useful for screening but also useful for tagging a person (suspect or normal) and assurance of his quarantine status. 

\begin{figure}
    \centering
    \includegraphics[scale= 0.4] {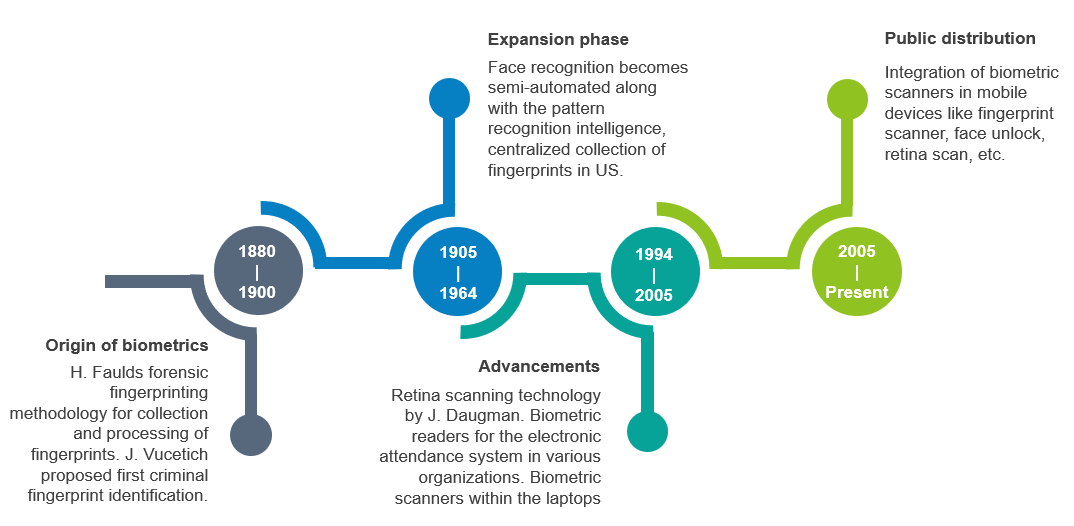}
    \caption{Evolution of biometrics over the years.}
    \label{fig25}
\end{figure}

Alphonse Bertillon, chief of the criminal identification division of the police department in Paris, developed and then practiced the idea of using a number of body measurements to identify criminals in the mid-19th century~\citep{jain2004introduction}. The term biometrics is derived from the greek words bio and metric; meaning life and measure respectively~\citep{barde2019classification}. The two main types of biometric identifiers depend on either physiological or behavioral characteristics used to digitally authenticate a person to access systems, devices or data. This is also useful for person tracking in automated surveillance systems. The basic premise of biometric authentication is that every person can be accurately identified by his or her intrinsic physical or behavioral traits. Physiological identifiers relate to the composition of the user being authenticated and include facial recognition, fingerprints, finger geometry (the size and position of fingers), iris recognition, vein recognition, retina scanning, voice recognition and DNA matching. Behavioral identifiers include the unique ways in which individuals act, including recognition of typing patterns, walking gait and other gestures. Some of these behavioral identifiers can be used to provide continuous authentication instead of a single one-off authentication check. In general, contact-based physiological biometrics such as fingerprint recognition pose a health risk~\citep{sharif2019overview}, hence contactless biometrics seem more promising such as facial and iris recognition. Fig.~\ref{fig25} shows the evolution in the biometric systems along with the survey papers in Table~\ref{tab15}.

\begin{table}[]
    \centering
    \caption{Survey papers on biometric systems.}
    \label{tab15}
    \resizebox{\linewidth}{!}{
    \begin{tabular}{|p{1.2in}|p{1.4in}|p{0.25in}|p{2.0in}|}
    \hline
        \textbf{Author}                     & \textbf{Title}                                                      & \textbf{Year} & \textbf{Contribution}                                                                                           \\
        \hline
\cite{kumari2019periocular} & Periocular biometrics: A survey                            & 2020 & Gives a deep insight of various aspects such as utility of periocular region as a stand-alone modality. \\  \hline
\cite{sharif2019overview}   & An Overview of Biometrics Methods                          & 2019 & Discussed all types of biometric systems.                                                              \\   \hline
\cite{gui2019survey}      & A Survey on Brain Biometrics                               & 2019 & Covers all aspects of brain biometrics.                                                                \\  \hline
\cite{gomez2020reversing} & Reversing the irreversible: A survey on inverse biometrics & 2019 & Discussed the importance of reverse engineering in the biometrics system.                              \\  \hline
    \end{tabular}}
\end{table}

\subsection{Analytics, storage and processing environment}

\subsubsection{Data analytics}

COVID-19’s infectious spread and severe impact in more than 190 countries have made the situation critical. The bad part of this pandemic over people is psychological weakness like: helplessness, scareness, and frustration. Data analytics methods are used to infer the patterns from available data to decide future actions. In presence of massive incremental data of this pandemic, advanced data analytics play a vital role. The analysis of the data may also help the people to know the tentative future behavioural patterns, that may surely become the best way to reduce panic and make people feel safe and hope in the tragic \enquote{pandemic} time. Data analysis helps health workers, administrators and researchers to prepare action plans in combat against COVID-19. Availability of COVID-19 related massive Data is an asset that is helping forecast and understands the reach and impact of coronavirus. In contrast to earlier epidemics, to fully exploit the potential of big data analytics approaches, the present advancements in computational hardware and resources are being used by healthcare workers, scientists and epidemiologists to analyze the data. 

\begin{figure}
    \centering
    \includegraphics[scale= 0.4] {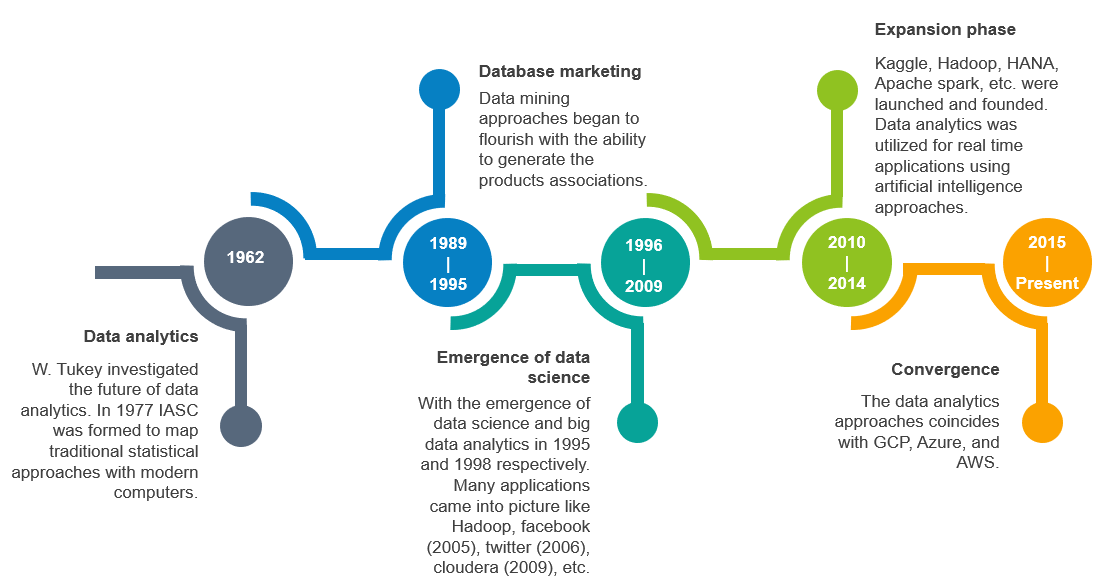}
    \caption{Breakthroughs in data analytics.}
    \label{fig11}
\end{figure}

The term data analytics is the backbone of every aspect of study that drives the approach towards its desired goal. Fig.~\ref{fig11} illustrates the advancements achieved over the decades in the field of data analytics. With the development of big data and advent of data sources, this field has significantly evolved to address the associated challenges of real-time processing. This area of research tends to discover the hidden patterns associated with the data by means of data-centric approaches like data mining, statistical analysis, data presentation and management~\citep{hu2014toward}. This mainly comprises of 2D2P analytics defined as descriptive, diagnostic, predictive, and prescriptive, where descriptive analytics answers what happened and past performance; diagnostic analytics answers why happened i.e. finds the cause of outcome of descriptive analytics; on the other hand predictive and prescriptive analytics answers what will happen in future and what should be done respectively~\citep{banerjee2013data}. Table~\ref{tab3} describes the recent surveys conducted to describe the emerging technologies in the area of data analytics.
\begin{table}[h!]
    \centering
    \caption{Survey papers on data analytics covering emerging technologies.}
    \label{tab3}
    \resizebox{\linewidth}{!}{
    \begin{tabular}{|p{1.2in}|p{1.4in}|p{0.25in}|p{2.0in}|}
    \hline
         \textbf{Author}                     & \textbf{Title}                                                                        & \textbf{Year} & \textbf{Main contribution}                                                                                               \\
         \hline
\cite{yurtsever2020survey}    & A survey of autonomous driving: Common practices and emerging technologies  & 2020 & Reviews the recent state-of-the-art approaches for automated driving systems by comparing the results in the common platform with real driving settings and environment.  \\ \hline
       \cite{nasiri2019evaluation}  & Evaluation of distributed stream processing frameworks for IoT applications in Smart Cities        & 2019 & Highlights the large-scale stream processing frameworks for high throughput and low latency applications of smart cities.    \\ \hline
\cite{inoubli2018experimental} & An experimental survey on big data frameworks                                                      & 2018 & Presents the best practices in big data frameworks to store, analyse and process the data.                                   \\ \hline
\cite{elijah2018overview}  & An overview of Internet of Things (IoT) and data analytics in agriculture: Benefits and challenges & 2018 & Explores the IoT ecosystem and data analytics concerned with smart agriculture.                                              \\ \hline
\cite{marjani2017big} & Big IoT data analytics: architecture, opportunities, and open research challenges                  & 2017 & Proposes new architecture model for big data IoT analytics and explores the relationship between IoT and big data analytics. \\ \hline
\cite{singh2015survey}    & A survey on platforms for big data analytics                                                       & 2015 & Addresses big data analytics  platforms with scalability, fault tolerance, I/O rate, etc.                                    \\ \hline
    \end{tabular}}
\end{table}

\subsubsection{Blockchain}
Since coronavirus is highly infectious in nature, people from all parts of the world are trying hard to get the appropriate solutions, in the form of fast detection of virus carriers and halting the spread of infection, and also in vaccine development. In the present context of epidemic management, blockchain is emerging as a crucial technology in order to provide a robust, transparent and low-priced method of enabling effective decision-making and, as an outcome, could lead to quicker actions during this crisis. Blockchain is now showing ample possibilities to become an integral part of the global response to COVID-19 by facilitating reliable track and trace applications for disease carriers, managing secure payments and maintaining the sustainable supply chain of essential goods and donations.

\begin{figure}
    \centering
    \includegraphics[scale= 0.4] {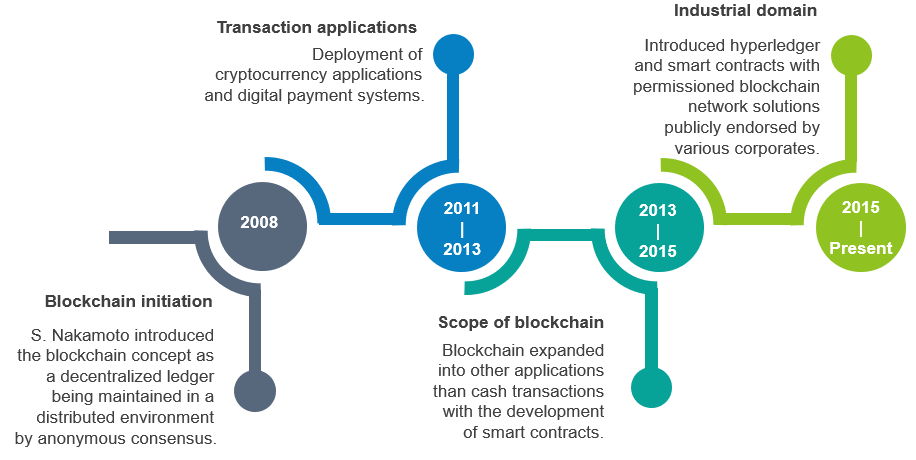}
    \caption{Development of blockchain scheme.}
    \label{fig12}
\end{figure}

Blockchain also known as the decentralized ledger, conceptualized by S. Nakomoto in 2008, is a cryptographic distributed database without any central controller~\citep{nakamoto2019bitcoin} that features cryptography, immutability, provenance, decentralization, anonymity, and transparency. As illustrated in Fig.~\ref{fig12} the concept of blockchain initiated as the electronic transaction system while expanding its dimensions in broad application areas e.g. cryptocurrency, banking, supply chain, property records, healthcare, security, privacy, etc.~\citep{pilkington2016blockchain}. The key components of blockchain network are node (user or computer within the blockchain), transaction (smallest building block of a blockchain system), block (a data structure used for keeping a set of transactions which is distributed to all nodes in the network), chain (a sequence of blocks in a specific order), miners (specific nodes which perform the block verification process), and consensus (a set of rules and arrangements to carry out blockchain operations). Table~\ref{tab4} describes the recent advancements in the blockchain technology with the help of survey papers.

\begin{table}[]
    \centering
    \caption{Recent surveys conducted in blockchain technology.}
    \label{tab4}
    \resizebox{\linewidth}{!}{
    \begin{tabular}{|p{1.2in}|p{1.4in}|p{0.25in}|p{2.0in}|}
        \hline
        \textbf{Author}                       & \textbf{Title}                                                                  & \textbf{Year} & \textbf{Main contribution}                                                                                                                                    \\
        \hline
\cite{maesa2020blockchain}     & Blockchain 3.0 applications survey    & 2020 & Other than cryptocurrencies, explores various application of blockchain in  healthcare records management, end-to-end verifiable electronic voting, access control systems, identity management systems, decentralized notary, etc.                                           \\ \hline
\cite{zubaydi2019review2} & A review on the role of blockchain technology in the healthcare domain & 2019 & Highlights the application of blockchain in the healthcare sector along with the generalized implementation framework.                               \\ \hline
\cite{panarello2018blockchain} & Blockchain and IoT integration: A systematic survey                    & 2018 & Presents the research trends in blockchain for IoT enabled environments and challenges associated with the IoT and blockchain.                       \\ \hline
\cite{chen2018survey}      & A survey of blockchain applications in different domains               & 2018 & Presents the recent advancements in blockchain in the broad areas of healthcare, insurance, energy, copyright protection, and societal applications. \\ \hline
\cite{joshi2018survey}    & A survey on security and privacy issues of blockchain technology       & 2018 & Presents an overview of the privacy and security concerns of the trends and applications in blockchain.
\\ \hline
\cite{dinh2018untangling}     & Untangling blockchain: A data processing view of blockchain systems    & 2018 & Proposed blockbench to analyse the performance of blockchain models for varying data processing workloads.                                           \\ \hline
\end{tabular}}
\end{table}

\subsubsection{Cloud computing}
The race for vaccine discovery of COVID-19 is underway, researchers and medical scientists across the world are working round the clock and sharing their research findings to the communities as an effort to bring the pandemic to an end. Lockdown and social distancing have triggered massive and sustained surge in users to stick to online platforms for sharing of work-related information. For all these reasons, cloud computing is facilitating video conferencing, remote project collaboration, e-education, etc. to meet spectacular and unplanned demand under this crisis.

\begin{figure}
    \centering
    \includegraphics[scale= 0.4] {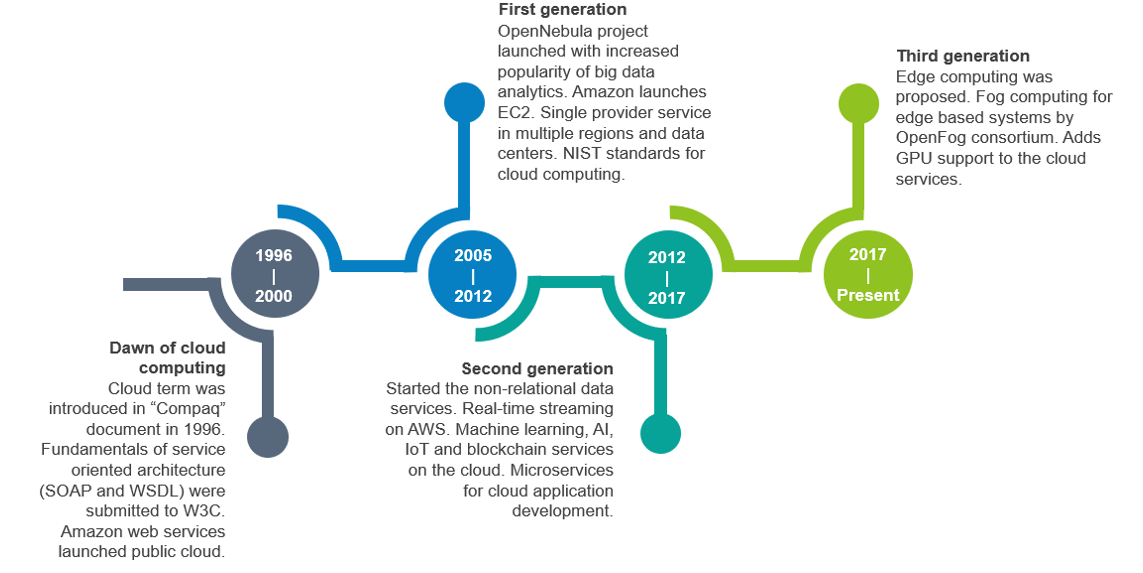}
    \caption{Timeline of event of emergence and development of cloud computing.}
    \label{fig18}
\end{figure}
\begin{figure}
    \centering
    \includegraphics[scale= 0.35] {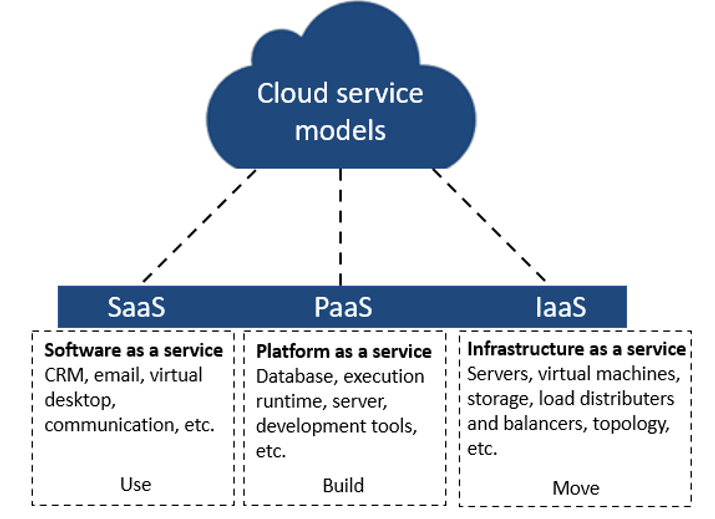}
    \caption{Different service models of cloud computing.}
    \label{fig19}
\end{figure}
\begin{table}
    \centering
    \caption{Recent surveys conducted in the field of cloud computing.}
    \label{tab9}
    \resizebox{\linewidth}{!}{
    \begin{tabular}{|p{1.2in}|p{1.4in}|p{0.25in}|p{2.0in}|}
    \hline
         \textbf{Author}                       & \textbf{Title}                                                                                                       & \textbf{Year} & \textbf{Main contribution}                                                                                                                                                                                        \\
         \hline
\cite{almusaylim2020comprehensive} & Comprehensive Review: Privacy Protection of User in Location-Aware Services of Mobile Cloud Computing & 2020 & Reviews the recent advances in the securing the privacy and location of the user in mobile cloud computing while also highlighting the shortcomings that need to be addressed.                                                 \\ \hline
\cite{arunarani2019task} & Task scheduling techniques in cloud computing: A literature survey                                          & 2019 & Highlights the exhaustive survey of task scheduling strategies for cloud environment along with its related issues, possible solutions and future scope.                                                 \\ \hline
\cite{nguyen2017virtual} & Virtual machine consolidation with multiple usage prediction for energy-efficient cloud data centers        & 2017 & Describes the energy efficient framework for cloud based computing based on virtual machine consolidation to estimate the long term resource utilization to mitigate the overload or burnout conditions. \\ \hline
\cite{madni2016resource}  & Resource scheduling for infrastructure as a service (IaaS) in cloud computing: Challenges and opportunities & 2016 & Present the resource scheduling schemes proposed so far for varying infrastructural environments along with the challenges and possible improvements.                                                    \\ \hline
\cite{singh2015qos}     & QoS-aware autonomic resource management in cloud computing: a systematic review                             & 2015 & Depicts the literature survey for efficient resource management and workload distribution in the cloud computational environment for best quality of offered services.                                   \\ \hline
\cite{mastelic2014cloud}  & Cloud computing: Survey on energy efficiency                                                                & 2014 & Proposes a systematic framework for energy efficient computations on cloud infrastructure along with the practice of state-of-the-art standard approaches.                                               \\ \hline
    \end{tabular}}
\end{table}

With the advancement in the above discussed technologies and data sources, the requirement of on-demand computation resources has increased. Cloud computing technology aims to provide the requirement based, on-demand high performance computing environment along with the data storage capability~\citep{voorsluys2011introduction}. The term was first coined in an internal document by \enquote{Compaq} in 1996~\citep{regalado2011coined}. The advancement in cloud computation follows three generations as shown in Fig.~\ref{fig18}, where the first generation offered a centralized network with two tier architecture, the second generation experienced the increase in the types of services and the number of service providers, and third generation improved the quality of resources by incorporation of GPU enabled accelerators. As per the NIST definition of cloud computing~\citep{mastelic2014cloud} services are offered in three different categories of model as shown in Fig.~\ref{fig19}. As per the survey observed (shown in Table~\ref{tab9}), the fourth generation of advancement in cloud is yet to come.

\subsubsection{Quantum computing}
It was seen in movies or studied in stories about an apocalyptic world where people use to fight against a deadly pathogen to find a cure. Unfortunately, this pandemic is not a fictional episode, it is actual, and experts across the globe are frantically observing for patterns in large sets of data by using powerful processors with the anticipation of finding a quicker breakthrough in vaccine/drug discovery for the COVID-19. Discovery of vaccine/drug may require screening of billions of chemical compounds quickly which can be efficiently possible with the hybridization of high performance computing and machine learning. 
\begin{figure}[H]
    \centering
    \includegraphics[scale= 0.4] {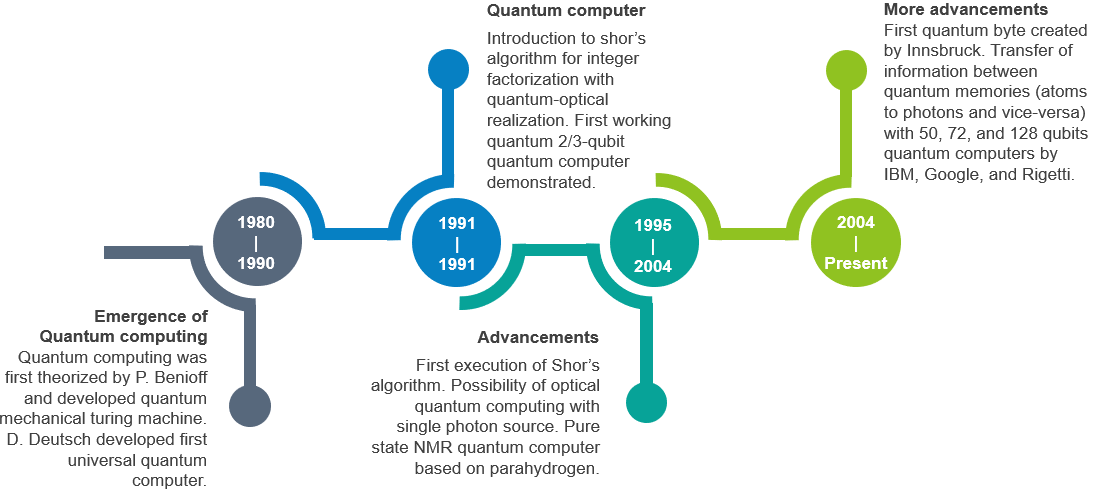}
    \caption{Development trends of quantum computing.}
    \label{fig24}
\end{figure}
Quantum computing brings classical information theory, computer science, and quantum physics together for faster computation compared to the conventional one~\citep{steane1998quantum}. This began with P. Benioff who was the first to theorize quantum computing in 1981~\citep{benioff1982quantum} and expanded into incredibly powerful machines that make use of the new approaches for processing information. Fig.~\ref{fig24} highlights the evolution of quantum computing over the years. Built on the principles of quantum mechanics, it exploits complex and fascinating laws of nature that are always there but usually remain hidden from view. By harnessing such natural behavior, quantum computing can run new types of algorithms to process information more holistically. It is believed that in the near future they may lead to revolutionary breakthroughs in materials and drug discovery, the optimization of complex man-made systems, and artificial intelligence. Quantum computing is expected to open the doors that were thought to remain locked indefinitely.  Table~\ref{tab14} has listed some important review articles on recent improvements in quantum computation.
\begin{table}[]
    \centering
    \caption{Literature surveys conducted in quantum computing.}
    \label{tab14}
    \resizebox{\linewidth}{!}{
    \begin{tabular}{|p{1.2in}|p{1.4in}|p{0.25in}|p{2.0in}|}
    \hline
         \textbf{Authors}                        & \textbf{Title}                                    & \textbf{Year} &  \textbf{Contribution}                                                                                                                                                                                        \\
         \hline
\cite{mcardle2020quantum}             & Quantum computational chemistry       & 2020                & Explores the approaches that can aid in building a quantum computer while also bridging knowledge gap between quantum computing and computational chemistry.                                                                                                                                  \\ \hline
\cite{gyongyosi2019survey}   & A Survey on quantum computing technology & 2019                & Discussed the most recent results of quantum computation technology and addressed the open problems of the field.                                                                                     \\ \hline
\cite{savchuk2019quantum} & Quantum Computing: Survey and Analysis   & 2019                & Discussed the main concepts and postulates of the quantum computing model, efficient quantum algorithms, and recent results, capabilities, and prospects in constructing a scalable quantum computer. \\ \hline
\cite{de2019quantum}         & Quantum Computing: Lecture Notes         & 2019                & Addressed the possible application domains, recent algorithms and unexplored area in the field of quantum computing.                                                                                  \\ \hline
\cite{schuld2019machine}             & Machine learning in quantum spaces       & 2019                & Talks about the feasibility of ML algorithms with quantum computers.                                                                                                                                  \\ \hline
    \end{tabular}}
\end{table}

\subsection{Autonomous transportation and communication technologies}

\subsubsection{Internet of things}
The entire globe is busy in fighting the menace of the COVID-19. At one hand, death tolls are shockingly rising and on the other hand new and innovative means of limiting the infection are widely being explored to stop the spread of the contagion. Since social distancing is continuously being given a push, all kinds of contactless solutions are receiving special attention. In this context, Internet of Things (IoT) is playing an important role in the technology arsenal because of its reduced cost, autonomous and remote caregiving features and easy diagnostic capabilities. In the pandemic response, smart connected devices can generate enough data for statistical analysis and subsequently aid in better decision making to gain prevention and control against COVID-19.

\begin{figure}
    \centering
    \includegraphics[scale= 0.4] {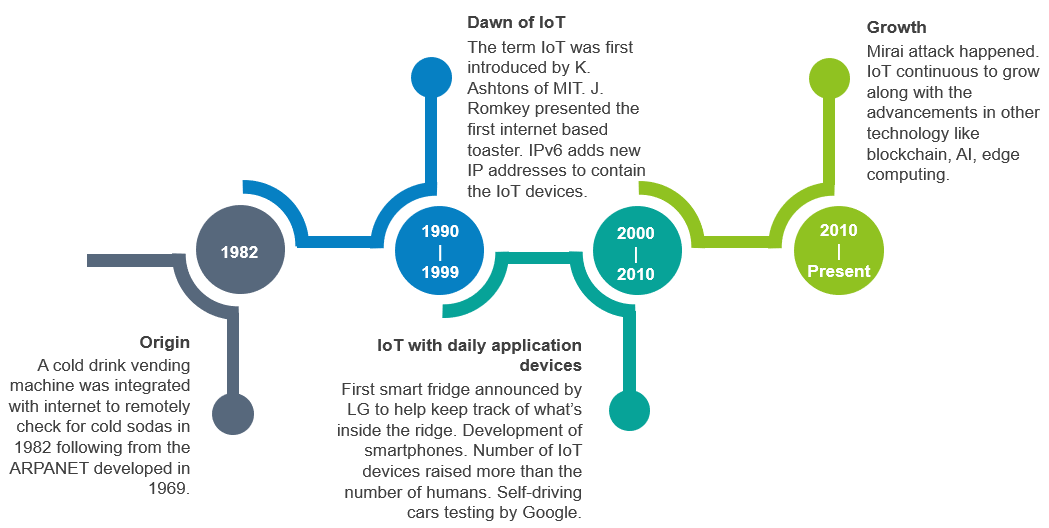}
    \caption{IoT timeline of advancement.}
    \label{fig16}
\end{figure}

The IoT corresponds to the web of interconnected physical devices, connected via internet, bluetooth or other means of connectivity~\citep{rayes2017internet}. The term IoT was first coined by Kevin Ashton that evolved from the smart \enquote{Coca Cola} vending machine in 1982 to smart home appliances, smart toys, smart healthcare, and a lot more due to the advancements in machine learning, artificial intelligence, real-time analytics, embedded systems, and other technologies. Fig.~\ref{fig16} represents the timeline of the IoT advancements since its origin. With its diverse application in real-world problems and innovations, it has been widely studied in integration with artificial intelligence techniques. Table~\ref{tab7} describes the recent literature surveys that exploits the application domain of IoT technology.
\begin{table}[]
    \centering
    \caption{Recent surveys conducted in the field of IoT.}
    \label{tab7}
    \resizebox{\linewidth}{!}{
    \begin{tabular}{|p{1.2in}|p{1.4in}|p{0.25in}|p{2.0in}|}
        \hline
         \textbf{Author}                       & \textbf{Title}                                                                                                  & \textbf{Year}& \textbf{Main contribution} \\
         \hline
\cite{younan2020challenges}     & Challenges and recommended technologies for the industrial internet of things: A comprehensive review         & 2020 & Surveys recent advances and recommended technologies in the IoT research and overview of information communication technologies along with its impact on the smart transportation and healthcare. \\ \hline
\cite{wang2019survey}     & Survey on blockchain for Internet of Things                                                            & 2019 & Elaborate on the existing blockchain technology from the IoT perspective along with the critical challenges that arise with their integration.                               \\ \hline
\cite{gharaibeh2017smart} & Smart cities: A survey on data management, security, and enabling technologies                         & 2017 & Explores the data centric perspective of IoT for smart cities to ensure consistency, interoperability, reusability and granularity of the acquired data.                     \\ \hline
\cite{alaba2017internet}   & Internet of Things security: A survey                                                                  & 2017 & Highlights the advancements in the security domain of the IoT in context of application, communication and architecture.                                                     \\ \hline
\cite{verma2017survey}   & A survey on network methodologies for real-time analytics of massive IoT data and open research issues & 2017 & Reviews the state-of-the-art IoT network analytics in real-time with different use cases and software platforms along with its respective shortcomings.                      \\ \hline
\cite{al2015internet} & Interne of things: A survey on enabling technologies, protocols, and applications                     & 2015 & Presents the in-depth overview of the IoT protocols and its relevance to the emerging technologies e.g. big data analytics, cloud computing, etc. and respective challenges. \\ \hline
\end{tabular}}
\end{table}

\subsubsection{Robotics}
Robots are one of the promising technologies supported during the fight against coronavirus. To contain the spread of coronavirus, robots are offering care to infected patients, quarantine people and managing social distancing. It is a time when robotics developers have to respond quickly to address the present need of public health concerns and in a most expedient and safest way so that the outbreak and its further spread can be grappled.
\begin{figure}
    \centering
    \includegraphics[scale= 0.4] {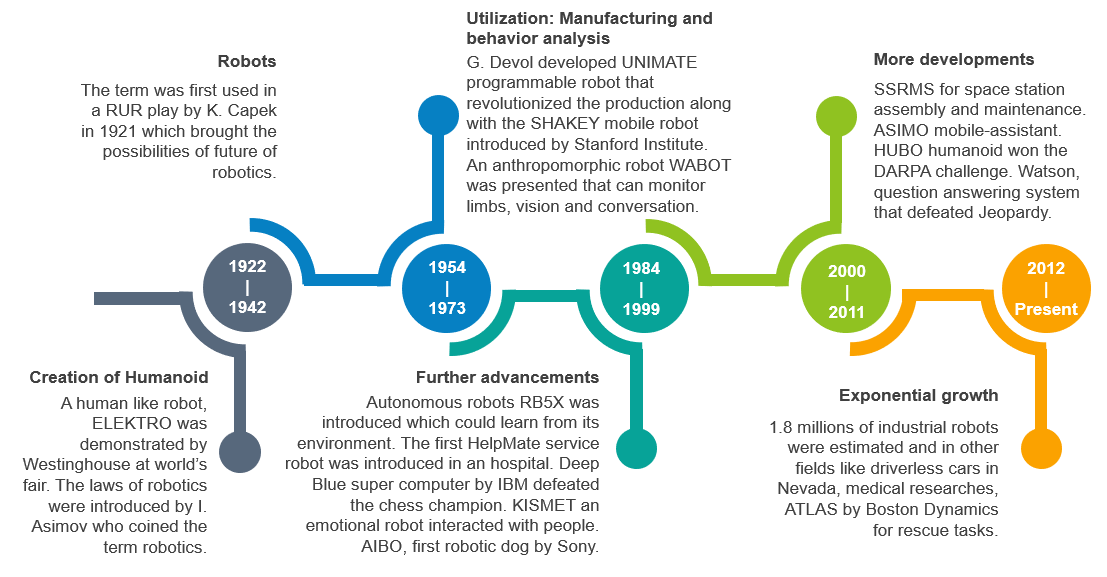}
    \caption{Advancements in the field of robotics.}
    \label{fig17}
\end{figure}
Robotics is an interdisciplinary field of study that covers physics, mathematics, mechanical engineering, computer science and engineering, dynamic system modelling, electrical and electronic engineering, etc.~\citep{najafi2018handbook}. This field studies to replicate the human actions using the automated self-learning robotic machines. The term was first coined in the play \enquote{Rossum's Universal Robots} or \enquote{R.U.R.} by the Czech writer Karel Capek. Starting from its first industrial application by J. F. Engelberger, as a robotic arm for die-casting, has widened the application area in various fields e.g. agriculture, automobile, entertainment, health-care, manufacturing, military, transportation, etc as shown in Fig.~\ref{fig17}. Table~\ref{tab8} indicates related works that show advancement in such areas. In this era, this area of research has faced major development with the advancements in artificial intelligence especially for creating robots that can replace humans. Many robots are built to do jobs that are hazardous to people such as defusing bombs, finding survivors in unstable ruins, and exploring mines and shipwrecks. Moreover, robotics is also used in STEM (science, technology, engineering, and mathematics) as a teaching aid. 

\begin{table}[h!]
    \centering
    \caption{Recent surveys conducted in the field of robotics.}
    \label{tab8}
    \resizebox{\linewidth}{!}{
    \begin{tabular}{|p{1.2in}|p{1.4in}|p{0.25in}|p{2.0in}|}
        \hline
         \textbf{Author}                        & \textbf{Title}                                                                                                                       & \textbf{Year} & \textbf{Main contribution}                                                                                                                               \\
         \hline
\cite{gualtieri2020emerging} & Emerging research fields in safety and ergonomics in industrial collaborative robotics: A systematic literature review & 2020 & Discusses the safety and ergonomics for industrial collaborative robotics with human–robot collaboration technology along with the emerging challenges and research area. \\ \hline
\cite{bing2018survey}   & A survey of robotics control based on learning-inspired spiking neural networks                                             & 2018 & Explores the advancements in the Spiking Neural Networks (SNNs) for the perspective of robotic applications.                                    \\ \hline
\cite{he2017survey}      & A survey of human-centered intelligent robots: issues and challenges                                                        & 2017 & Describes the approaches and mlkchallenges for the advancements in the development of human centric robots.                                      \\ \hline
\cite{polydoros2017survey} & Survey of model-based reinforcement learning: Applications on robotics                                                      & 2017 & Incorporates the proposed approaches on model-based reinforcement learning with its applicability in the area of robotics.                      \\ \hline
\cite{eguchi2016robocupjunior}    & RoboCupJunior for promoting STEM education, 21st century skills, and technological advancement through robotics competition & 2016 & Highlights the case study of RoboCupJunior in relation with STEM based learning for creative and innovative learning.                           \\ \hline
\cite{kostavelis2015semantic} & Semantic mapping for mobile robotics tasks: A survey                                                                        & 2015 & Presents the approaches to bridge the gap between humans and robots interpretation of information by incorporating semantic mapping techniques. \\ \hline
    \end{tabular}}
\end{table}

\subsubsection{5G mobile technologies}
5G is the fifth generation of wireless communications technologies which is 10 times faster than fourth-generation technology. Presently, smartphones are very common among people for voice calls, social networking activities, online banking, telemedicine, etc. COVID-19 is at its peak and seeking to maintain social distancing as the most prominent solution for stopping the spread of the virus, people are shifting towards fully online mode like e-learning, conferences, meetings, shoppings, telemedicine, etc. As the number of active internet users are drastically increased, existing mobile networks are not sufficient to cope up the situation. Hence, 5G technology is the primary requirement to fulfill the current and future requirements.

\begin{figure}
    \centering
    \includegraphics[scale= 0.4] {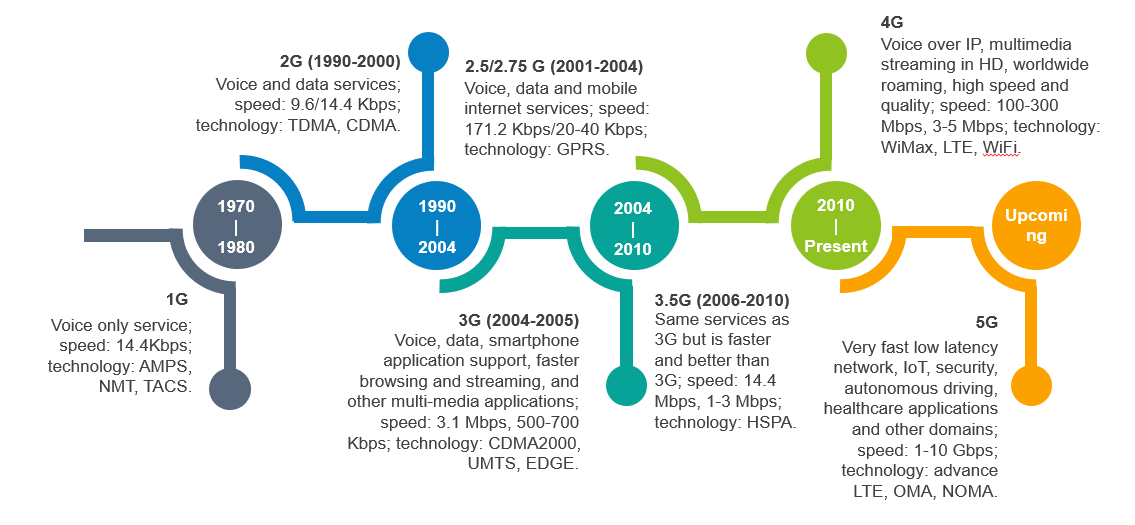}
    \caption{Evolution of wireless generation of telecommunication technologies.}
    \label{fig20}
\end{figure}

Unlike its predecessor, 5G offers greater bandwidth and higher download speed that can scale up to 10 gigabits per second (Gbps). Fig.~\ref{fig20} highlights the evolution of wireless technologies~\citep{sharma2013evolution}. With increased bandwidth in 5G, it can compete with other internet service providers (ISP) like cable net that is utilized in laptop or desktop devices and can also widen the scope of applications in IoT and machine-to-machine (M2M) enabled regions. This quality of telecommunication can serve as a foundation of many aforementioned technologies. 5G is now recognized as a disruptive technology with its forthcoming deployment between 2020 and 2040 across the world. Table~\ref{tab10} highlights the literature surveys conducted for 5G and related telecommunication technologies. Since networking and mobile connectivity is the base for every technology, the implication of 5G will directly impact other technologies in terms of data access and transfer rates.

\begin{table}[h!]
    \centering
    \caption{Recent surveys conducted in the field of 5G telecommunications.}
    \label{tab10}
    \resizebox{\linewidth}{!}{
    \begin{tabular}{|p{1.2in}|p{1.4in}|p{0.25in}|p{2.0in}|}
    \hline
         \textbf{Author}                    & \textbf{Title}                                                                                                       & \textbf{Year} & \textbf{Main contribution}                                                                                                                                                             \\
         \hline
\cite{mistry2020blockchain}  & Blockchain for 5G-enabled IoT for industrial automation: A systematic review, solutions, and challenges                   & 2020 & Provides the comprehensive review on blockchain based 5G enabled IoT approach along with its industrial potential while also addressing the associated challenges and future research possibilities.  \\  \hline
\cite{su2019resource}   & Resource allocation for network slicing in 5G telecommunication networks: A survey of principles and models & 2019 & Describes the resource allocation schemes in 5G networks to highlight load balancing, resource utilization, and network performance.                                          \\  \hline
\cite{shaikh2019comprehensive} & Comprehensive Survey of Massive MIMO for 5G Communications                                                  & 2019 & Presents the integration of MIMO concepts with the 5G networking along with the best suited mili-meter wave spectrum this telecommunication.                                   \\  \hline
\cite{agiwal2016next} & Next generation 5G wireless networks: A comprehensive survey                                                & 2016 & Presents the exhaustive survey of all the related technologies and advancements for 5G communication along with its evolution and challenges.                                 \\  \hline
\cite{zou2016survey}  & A survey on wireless security: Technical challenges, recent advances, and future trends                     & 2016 & Highlights the security, vulnerabilities and threat aspects arising with the advancements in the wireless telecommunication.                                                   \\  \hline
\cite{yu20113g} & From 3G to 4G: technology evolution and path dynamics in China's mobile telecommunication sector            & 2011 & Investigates the technologies or advancements that drove this evolution in telecommunication along with the practices of different stakeholders with the case study of China. \\  \hline
    \end{tabular}}
\end{table}

\subsubsection{Autonomous vehicle}
During this global crisis of COVID-19 pandemic, to restrict the spread of the infectious coronavirus (by means of reducing people-to-people contact), autonomous vehicles are surely the best transportation medium to deliver essential goods like medicines and food items. Autonomous vehicles are also useful in several areas during this pandemic like: isolation wards in hospitals, railway stations, airports, etc. The first witnessed application of autonomous vehicles during this pandemic was accomplished by Apollo (a Baidu’s autonomous vehicle platform) to supply daily needs, medicines and food to a hospital at Beijing.
\begin{figure}[H]
    \centering
    \includegraphics[scale= 0.4] {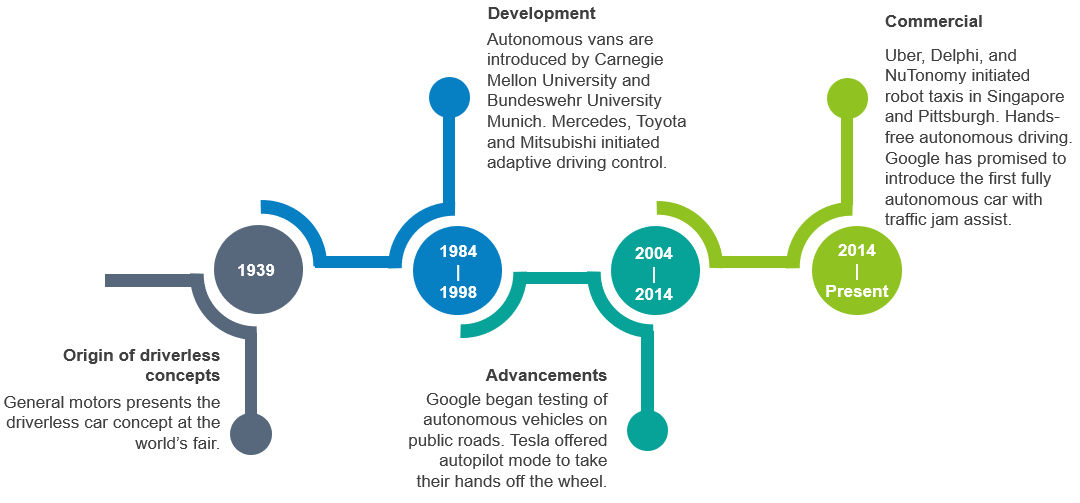}
    \caption{Timeline of Autonomous vehicles.}
    \label{fig26}
\end{figure}
Autonomous vehicles~\citep{yurtsever2020survey} are one of the most desirable future transportation techniques. The evolution of autonomous vehicles follows from the objective to develop these vehicles is to take  over the art of driving by computing devices for more promising transportation. These vehicles can drive themselves from source to destination in “autopilot” mode using various in-vehicle technologies and sensors such as adaptive collision control, active steering (steer by wire), anti-lock braking systems (brake by wire), GPS navigation technology, lasers, etc. An autonomous car is self-guided without any human conduction. Google’s autonomous car design, have logged thousands of hours on American roads, but they are not yet commercially available on a large scale~\citep{fagnant2015preparing}. Studies suggest that significant autonomous car production could cause problems with existing auto insurance and traffic controls used for human-controlled cars~\citep{bimbraw2015autonomous}. Continuous  efforts for promising and robust autonomous vehicles are underway across the globe. It is only a matter of time before these kinds of advances allow us to outsource our daily commute to a computer. Table~\ref{tab16} gives a brief overview of development in autonomous vehicles and Fig.~\ref{fig26} shows timeline of automobile evolution.

\begin{table}[]
    \centering
    \caption{Overview of Autonomous vehicles.}
    \label{tab16}
    \resizebox{\linewidth}{!}{
    \begin{tabular}{|p{1.2in}|p{1.4in}|p{0.25in}|p{2.0in}|}
    \hline
         \textbf{Author}                    & \textbf{Title}                                                                                                                         & \textbf{Year} & \textbf{Contribution}                                                                                                                    \\
         \hline
\cite{yurtsever2020survey} & A survey of autonomous driving: common practices and emerging technologies                                                    & 2020 & Discussed the feasibility of existing automated driving systems (ADSs) for a safe, comfortable and efficient driving experience. \\ \hline
\cite{liu2020public}      & Public attitude toward self-driving vehicles on public roads: Direct experience changed ambivalent people to be more positive & 2020 & Discussed the resistance and negative attitudes to self-driving vehicles.                                                        \\ \hline
\cite{badue2019self} & Driving Cars: A Survey                                                                           & 2019 & The overall architecture is organized into the perception system and the decision-making system.        \\ \hline
\cite{dixon2020drives}     & What drives support for self-driving car technology in the United States?                                                     & 2018 & Favours the self driving over traditional human driven cars.                                                                    \\ \hline
    \end{tabular}}
\end{table}

\subsubsection{Drone technologies}
During the lockdown, when social distancing is of utmost priority, drones are being used on the front line to contain the virus spread. It is supporting disinfection tasks, patrolling activities, surveillancing and automatic delivery agents of food and other essentials in quarantined districts. These special needs of worldwide quarantine exercises are becoming a driving factor in the drone development sector leading to develop special purpose drones for specific tasks such as to help law and order management through automated surveillance.

\begin{figure}
    \centering
    \includegraphics[scale= 0.45] {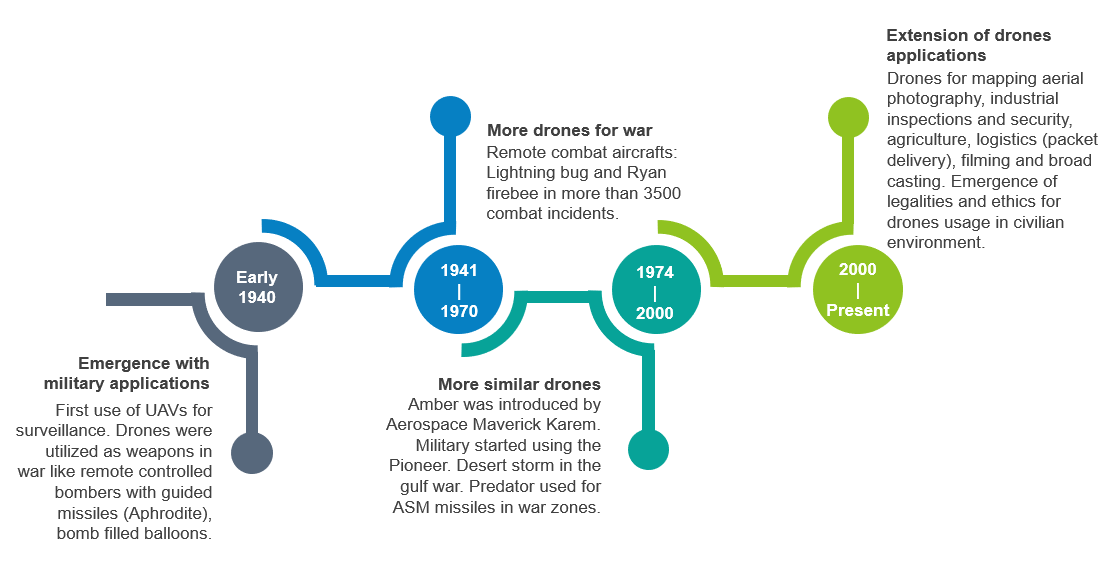}
    \caption{Evolution of drone technology.}
    \label{fig21}
\end{figure}

An uncrewed aerial vehicle (UAV) (generally known as a drone) is an unmanned aircraft~\citep{gupta2013review} initially, ~\citep{newcome2004unmanned}. It contains three components: UAV, ground control station and communication medium. The drones can be differentiated in terms of the type (fixed-wing, multirotor, etc.), the degree of autonomy, the size and weight, and the power source~\citep{vergouw2016drone}. Probably, the basic properties of UAVs will remain unchanged in future, but the aspects like propulsion, autonomy, and size may vary. The drone types are mainly categorized as fixed-wing, multirotor and hybrid systems (ornithopters). A fixed-wing drone uses fixed static wings in combination with forwarding airspeed to generate lift like traditional airplanes. Multirotor systems are a subset of rotorcraft  (defined as an aircraft that uses rotary wings to generate lift) e.g. traditional helicopters. Drones use multiple small rotors, whereas hybrid systems incorporate properties of both fixed-wing and multirotor systems. A hybrid quadcopter is a perfect example of a hybrid drone. This type of drone uses multiple rotors to lift and land, whereas wings are helpful to fly for longer distances. Fig.~\ref{fig21} shows the incremental growth in drone technology whereas Table~\ref{tab11} incorporates several articles on the state-of-the-art of drone technology.

\begin{table}[h!]
    \centering
    \caption{Recent surveys conducted in the field of drone technology.}
    \label{tab11}
    \resizebox{\linewidth}{!}{
    \begin{tabular}{|p{1.2in}|p{1.4in}|p{0.25in}|p{2.0in}|}
    \hline
         \textbf{Author}                    & \textbf{Title}                                                                                                                         & \textbf{Year} & \textbf{Contribution}                                                                                                                    \\ \hline
\cite{sharma2020communication}    & Communication and networking technologies for UAVs: A survey  & 2020 & Discusses the insight of emerging communication technologies for unanmed aerial vehicles along with its integration with the recent machine learning approaches to meet the modern needs of surveillance, navigation and edge computing. \\ \hline
\cite{kugler2019real}          & Real-world applications for drones                                                                 & 2019 & Talks about various application domains where drones can be applied. \\ \hline
\cite{li2019applications}    & Applications of multirotor drone technologies in construction management                           & 2019 & Discussed all important issues and applicability of drone technology. \\ \hline
\cite{vergouw2016drone} & Drone technology: Types, payloads, applications, frequency spectrum issues and future developments & 2016 & Talks about type and application areas of   drone technology.        \\ \hline
\cite{gupta2013review}   & Review of unmanned aircraft system (UAS)                                                           & 2013 & State-of-the-art of drone technology.                                \\ \hline
    \end{tabular} }
\end{table}

\section{Different ways to use disruptive and emerging technologies against COVID-19}
Preceding section offers a brief overview of the technologies currently in use, this section highlights their foremost features and significance in the fight against the coronavirus pandemic, converging on the methods of their utilization to monitor and contain the rapid spread of the virus, and to sensitize the authorities to develop and maintain their capacity toward fulfilling  the ever-increasing needs triggered by this pandemic. Recent research works related to various use cases concerning  COVID-19 are presented in the following subsections:

\subsection{Prediction and propagation analysis}
The objective of prediction follows from the idea of analyzing the past experience to forecast the upcoming possibilities. To fulfil this objective, AI-based approaches along with the assistance of other technologies e.g. cloud computing, IoT, etc., are widely studied. With the uncontrolled, non-stop exponential growth of the COVID-19 cases, many factors or parameters that drive the world are impacted in terms of the global economy~\citep{mckibbin2020global}, medical resources~\citep{emanuel2020fair}, business value~\citep{bartik2020small}, climate conditions~\citep{wang2020high}, etc. In order to analyse the impact of COVID-19 on such parameters, researchers use official data collected from publicly available datasets~\citep{kaggle} to analyse the propagation factor of the COVID-19 via prediction of the possible number of new cases, based on the susceptible population vulnerable to the exposure of a specific region or across the world~\citep{punn2020covid}. Furthermore, following the same concept and statistical analysis, the peak infection rates can be computed for a certain set of conditions imposed from the proposed hypothesis. Based on these findings, the table~\ref{tab17} highlights the SWOT analysis performed on these works.

\begin{table}[]
    \centering
    \caption{SWOT analysis in prediction and propagation analysis of COVID-19.}
    \label{tab17}
    \resizebox{\linewidth}{!}{
    \begin{tabular}{|p{0.8in}|p{5.0in}|}
    \hline
        \textbf{SWOT} & \textbf{Remarks}\\ \hline
    
         Strength      & Quick/time-saving analysis of the COVID-19 related parameters without manual intervention.                                                                                                                               \\ \hline
Weakness      & Rely heavily on the patterns of the already identified COVID-19 cases which may generate the biasness concerned with the incoming drifting data.                                                                         \\ \hline
Opportunities & The predictive analysis will give a chance to minimize the amount of loss that may occur based upon the current situation of COVID-19 by generating the follow-up plans and mitigate the issues with the early planning. \\ \hline
Threats       & With the possibility of false-negative predictions, an unnecessary panic situation could be generated. Security concerns related to the private information in the globally shared data.                                 \\ \hline
    \end{tabular}}
\end{table}
\subsection{Detection: diagnosis and prognosis}
AI has been playing an essential role in the areas of medical image processing, object detection, text mining, natural language processing, IoT, etc.~\citep{punn2020inception, zhang2018deep, young2018recent, punn2019crowd}. In detection, an object is identified to aid in the underlying processing of the data. AI-based approaches including deep learning and machine learning are explored by young researchers and laureates to derive measures that could aid in early detection of COVID-19 by utilizing the data from various sources consisting of medical images, statistical IoT data, news and reports, etc.

In the study of medical image processing via deep learning approaches, many works have been proposed that make use of chest X-ray and CT scan images for the detection of COVID-19 at the earliest possible stage~\citep{punn2020automated, wong2020frequency, li2020coronavirus, narin2020automatic, punn2020chs} . This reduces the overhead time of radiologists spent into the close monitoring of these images for further diagnosis. Furthermore, till now, no concrete solution is proposed to fight against COVID-19 and hence social distancing~\citep{greenstone2020does} is found to be the only feasible solution. Deep learning based object detection and tracking schemes are proposed, followed by statistical analysis to automate the monitoring of social distancing via surveillance footage~\citep{punn2020monitoring}. Likewise, many frameworks are proposed that make use of sensors available on the smartphones and other wearable devices like GPS, camera, temperature, heart rate monitor, and other inertia based sensors~\citep{maghdid2020novel}, to carefully monitor the victims of COVID-19. The data acquired is then utilized with machine learning approaches to synthesize the symptoms of the virus and later use it for its faster detection. This is so-far the quickest approach to detect COVID-19 as compared to the medical scans or kits, but is prone to false detections. With the help of above discussed works, table~\ref{tab18} indicates the SWOT analysis performed in the detection area of fight against COVID-19.

\begin{table}[]
    \centering
    \caption{SWOT analysis of detection task of COVID-19.}
    \label{tab18}
    \resizebox{\linewidth}{!}{
    \begin{tabular}{|p{0.8in}|p{5.0in}|}
    \hline
        \textbf{SWOT} & \textbf{Remarks}\\
        \hline
        
         Strength      & On-time or early detection of COVID-19 resulting in faster cure and treatment.                                                                                                                                                                                                                                                                  \\ \hline
Weakness      & Proposed AI approaches require heavy computation resources for the detection of COVID-19 which might not be available everywhere.                                                                                                                                                                                                               \\ \hline
Opportunities & AI enabled techniques support the diagnostic procedure with automated approaches to detect symptoms pertaining to COVID-19 with the help of medical modalities, IoT, drug discovery, etc. resulting in faster discovery of the virus. More time can be devoted for cure and treatment instead of manual monitoring of the symptoms of COVID-19. \\ \hline
Threats       & The AI approaches are not 100\% reliable and hence false detection of the COVID-19 could result in backfire, thereby making manual supervision of the results necessary. Medical data contains the private and crucial information of the patients and hence should be secured before it can be shared in the public platform.                  \\ \hline
    \end{tabular}}
\end{table}
\subsection{Treatment and drug discovery}
AI based approaches are not only focused on analysing the medical data to assist in the diagnosis practice but are also suitable for drug discovery to cure COVID-19. Meanwhile, in the on-groing research to understand the protein structures associated with COVID-19, Google’s AlphaFold~\citep{senior2020improved} model is being utilized to understand how COVID-19 operates and what are the possibilities of therapeutic drug discovery.~\cite{zhavoronkov2020potential} utilized deep learning based generative approaches e.g. generative autoencoders, genetic algorithm, generative adversarial network, etc., to design protease inhibitors to contain COVID-19. The proposed approach follows a reinforcement learning scheme to construct complex structures of molecules that can counter the effects of COVID-19. At the same time,~\cite{ton2020rapid} proposed deep learning based deep docking (DD) framework for rapid identification of fine inhibitors of COVID-19, allowing faster computation of docking scores for billions of plausible molecules with virtual screening. Similarly,~\cite{randhawa2020machine} aimed to classify the novel COVID-19 pathogens using their genomic structure with the help of alignment-free whole-genome and decision tree approach. With extensive trials conducted on over 5000 viral samples, the proposed hypothesis was found to fit with the bat origin of COVID-19 as claimed by~\cite{zhou2020pneumonia}. With these on-going researches and advancements, Table~\ref{tab19} describes the SWOT analysis.
\begin{table}[]
    \centering
    \caption{SWOT analysis of computational drug discovery and treatment of COVID-19.}
    \label{tab19}
    \resizebox{\linewidth}{!}{
    \begin{tabular}{|p{0.8in}|p{4.0in}|}
    \hline
        \textbf{SWOT} & \textbf{Remarks}\\
        \hline
         Strength      & Cost effective and time efficient approach for the COVID-19 treatment discovery.                                                                                                                           \\ \hline
Weakness      & Wrong hypothesis discovery may generate fatal results in the real world.                                                                                                                                   \\ \hline
Opportunities & AI based approaches for treatment discovery can be executed without causing any harm to the living society.                                                                                                \\ \hline
Threats       & The artificial simulations of treatment discovery may deviate from the real world and hence could cause loss of lives. Risk of privacy breach of the patient’s private information on the public platform. \\
\hline
    \end{tabular}}
\end{table}
\subsection{Telemedecine and assistance on demand}
The rampant of COVID-19 cases has brought global crisis and shortage of healthcare practitioners around the world. To mitigate this issue and reduce the workload, AI based automated healthcare assistance in terms of chatbots ~\citep{NP82}, self assessment applications~\citep{NP83}, AI assisted phone calls~\citep{NP84} and other expert systems~\citep{NP85}, are developed which would help to increase the reachability of necessary information concerned with the outbreak, precautionary measures, symptoms, decisive actions, etc. Dedicated chatbot for COVID-19 is so far the quickest approach possible to resolve the respective concern of a person, AI assisted phone calls can also do the same job with real-time streaming of audio messages, whereas self assessment applications make use of certain set of questionnaire and rankings by which it declares the status of symptoms being positive or negative for COVID-19. For uncertain situations where more supervision or guidance is required people can also make use of applications like FreeStyle Libre~\citep{NP86} based on cloud technology which can remotely connect the person with the doctor for distant monitoring of the patient. As for another approach people can share their respective data acquired from equipped sensors (smartwatches, smartphones) like heart rate, blood pressure on the global platform so that quick and immediate actions can be taken for the concerned person. Table~\ref{tab20} presents the SWOT analysis of the above discussed approaches.

\begin{table}[]
    \centering
    \caption{SWOT analysis of the scope of telemedicine during COVID-19 pandemic.}
    \label{tab20}
    \resizebox{\linewidth}{!}{
    \begin{tabular}{|p{0.8in}|p{4.0in}|}     \hline      \textbf{SWOT} & \textbf{Remarks}\\         \hline
        Strength      & Quick and cost effective approach for self diagnosis. Reduces the workload of the hospitals and other needed resources. Generate self awareness about the COVID-19 in the society while staying at home.                                                                                                                       \\ \hline
Weakness      & Require good quality of resources which may not be available in remote locations. Spread of false information would generate unnecessary panic situations.                                                                                                                                                                     \\ \hline
Opportunities & Gives a chance to the experts to concentrate more on severe cases for cure, treatment, and drug discovery.                                                                                                                                                                                                                     \\ \hline
Threats       & This type of diagnosis may raise the anxiety level among people which may result in the execution of unwanted actions. People may share  false information on public platforms which could  mislead the community. Security and privacy is also the matter of concern for the people sharing their vital information globally. \\ \hline
    \end{tabular}}
\end{table}
\subsection{Spread containment}
COVID-19 transmits from person-to-person via direct/indirect contact from the infected person~\citep{NP94}, the spread confinement measures are required to be adopted. The technical advancements have contributed in this area by releasing various technology driven contactless initiatives~\citep{NP98} e.g. rapid testing kits~\citep{NP87}, sanitization tunnels~\citep{kwon2020drive}, face shields~\citep{livingston2020sourcing}, bubble helmet~\citep{biffi2020considerations}, smart sanitization and screening systems~\citep{wosik2020telehealth}, personal protective equipment (PPE) kits~\citep{livingston2020sourcing}, thermal cameras~\citep{world2020rational}, POD units~\citep{kwon2020drive}, etc.

Rapid testing kit is a cost effective approach that offers assistance to the healthcare practitioners in faster diagnosis of the COVID-19 with triaging patients. The sanitization tunnels that act as disinfection spraying channels along with the thermal supervision to monitor the body temperature, ensure the hygiene maintenance of the person who was exposed to the outside environment. Though this approach is not cost effective but is necessary to sustain the hygienic surrounding in the public environment e.g. airport, malls, markets, etc. To maintain self hygiene, face shields, PPE kits and smart sanitization systems are introduced that make use of IoT devices to guide a person to follow the suggested measures properly; for instance, washing hands for 20 seconds, wearing a mask and other protective gears while going outside from home, following the social distancing order, etc. Recently, various cabinet structures like PODs are introduced to safeguard a person from getting in contact with the other in public places e.g. restaurants in Amsterdam are equipped with quarantine greenhouses~\citep{NP96}, isolation pods are installed in hospitals and airports, etc. Moreover, various hospitals are now equipped with bubble helmet based ventilation schemes, which is highly recommended for a patient suffering from severe respiratory syndrome than just by mere face mask~\citep{feng2020rational}. Following from these spread containment measures, table~\ref{tab22} presents their SWOT analysis.

\begin{table}[]
    \centering
    \caption{SWOT analysis of spread containment measures of COVID-19.}
    \label{tab22}
    \resizebox{\linewidth}{!}{
    \begin{tabular}{|p{0.8in}|p{5.0in}|}     \hline      \textbf{SWOT} & \textbf{Remarks}\\         \hline
         
Strength      & Cost effective measure to contain the virus spread.                                                                      \\ \hline
Weakness      & The required resources may not be available world wide. Equipment too needs sanitization and supervision from an expert. \\ \hline
Opportunities & Encourages to develop healthy habits with self care and self hygiene.                                                    \\ \hline
Threats       & Equipments need to be sterilized after every use, else it would be disastrous.                                            \\ \hline
    \end{tabular}}
\end{table}
\subsection{Fighting \enquote{infodemic}}
There is a growing concern about propagation of false and misleading information about the COVID-19 pandemic appearing in public discourse, especially on social media. Organisations like WHO and UN are taking initiatives and launching platforms to combat misinformation around COVID-19~\citep{SA21, SA22}. Trending techniques such as AI can be exploited to develop policies and guidelines about the fair use of all such platforms, for culling fake news and rumors, thus curbing the \enquote{infodemic} to a great extent~\citep{zarocostas2020fight}. There should be genuine and authentic availability of streamline search engines, so that search queries should get routed to the authentic and scientific information. In addition, AI chatbots can be used to propagate correct information about the disease and its symptoms under the strict supervision of competent authorities such as centers for disease control and prevention (CDCP). There are many regulatory organizations working in collaboration with various social media platforms to limit the spread of rumours, which may contain misrepresentation of various facts such as the virus will automatically vanish in hot seasons, consuming chloroquine can cure the coronavirus infection, and consuming large amounts of ginger and garlic can prevent the virus. The infodemic prevention techniques include aggressively filtering out groundless medical advice, hoaxes and other incorrect information that are fatal to the public health~\citep{cinelli2020covid}. In table~\ref{tab23}, SWOT analysis for the above discussed solutions for preventing  social infodemic is presented.

\begin{table}[]
    \centering
    \caption{SWOT analysis of the technology based solutions in prevention of social infodemic.}
    \label{tab23}
    \resizebox{\linewidth}{!}{
    \begin{tabular}{|p{0.8in}|p{5.0in}|}     \hline      \textbf{SWOT} & \textbf{Remarks}\\         \hline
        
Strength      & Reduces rumors concerning COVID-19 with automated approaches using artificial intelligence schemes.  \\ \hline
Weakness      & Intensive natural language processing is needed to identify fake news.                               \\ \hline
Opportunities & Better learning algorithms can be implemented for more robust solutions to prevent social infodemic. \\ \hline
Threats       & False rejection is possible even if the floated information is of public concern.                    \\ \hline
    \end{tabular}}
\end{table}

\subsection{Discovery}
With the discovered hypothesis of the origin of COVID-19 from bat~\citep{rothan2020epidemiology}, the person experiencing the COVID-19 shows the following symptoms being monitored with the help of data analytics techniques over the course of disease: fever (83–99\%), cough (59–82\%), fatigue (44–70\%), anorexia (40–84\%), shortness of breath (31–40\%), sputum production (28–33\%), and myalgias (11–35\%)~\citep{NP105}. Currently, most of the clinical measures on COVID-19 pandemic are focused on infection prevention, containment measures, and supportive cares for supplementary oxygen and ventilatory support including immunity enhancements; meanwhile,  several trials are conducted for the development of concrete solutions to counter COVID-19 which can be found at ClinicalTrials~\citep{NP100}. Following the current situation, the national institute of health (NIH) proposed general guidelines~\citep{NP101} for clinicians to assist in the treatment of COVID-19 with the use of data analytics and AI enabled approaches. The data is regularly updated based on the discoveries made by the authorized departments or from the published data. As per the current guidelines, recommended treatments are proposed for infection control, hemodynamic support, ventilatory support and drug therapy. Apart from these practices, the Ayurveda department has also begun its interventions as indicated by~\cite{rastogi2020covid} while exploiting its potential for the possibilities in cure and treatment of COVID-19 with the elaborative description of relevant epidemic conventional measures as mentioned by~\cite{jyotirmoy2016concept}. Besides these many efforts, till now, no drug has proven to be effective in successful treatment of COVID-19. Table~\ref{tab24} highlights the SWOT analysis of the above discussed findings.
\begin{table}[]
    \centering
    \caption{SWOT analysis of discovery approaches for COVID-19.}
    \label{tab24}
    \resizebox{\linewidth}{!}{
    \begin{tabular}{|p{0.8in}|p{4.0in}|}     \hline      \textbf{SWOT} & \textbf{Remarks}\\         \hline
         
Strength      & May lead to discover permanent cure for COVID-19.                                                   \\ \hline
Weakness      & Several trials would consume a huge amount of resources, efforts, time and financial support.       \\ \hline
Opportunities & Discovery of novel cure, treatment and monitoring  procedures along with the healthcare equipments. \\ \hline
Threats       & Novel discovered drugs may generate severe side effects.                                            \\ \hline
    \end{tabular}}
\end{table}
\subsection{Preparedness to economic disruption}
The COVID-19 pandemic has brought global crisis with the steep decay in many areas or sectors like travel, events (sports, concerts, conference, etc.) and business (import-export, construction, market, etc.), due to the economic consequences of boundations and lockdown. If this pandemic continues then GDP growth could fall to zero~\citep{NP108}. However, it is believed that technological advancements withhold the potential to balance the nation’s economy. 

According to the survey of Accenture~\citep{NP106}; by 2035 it is estimated that technological advancements and applications would contribute to balance the global economy. Whereas NASSCOM~\citep{NP107} reported that there has been a steep decline in the growth rate of IT based organizations. With almost every IT firm switching to work from home schemes with remote access, employees' full potential can be utilized in developing future intensive applications to meet the demands of the post COVID-19 world~\citep{NP109}. In this era of pandemic, the AI enabled applications can boost up the GDP growth with help of analysing the demand trends using data analytics and sales forecasting, chatbots to offer assistance and recommendation to the customers, integration of items availability in the nearby stores with google maps, and digital marketing with big data analytics. With these many efforts we can mitigate the impact of COVID-19 on the global economy. Table~\ref{tab25} describes the SWOT analysis of the technology based advancements to prepare for economic disruption.

\begin{table}[]
    \centering
    \caption{SWOT analysis of the technology based preparedness of economic disruption.}
    \label{tab25}
    \resizebox{\linewidth}{!}{
    \begin{tabular}{|p{0.8in}|p{4.5in}|}     \hline      \textbf{SWOT} & \textbf{Remarks}\\         \hline
Strength      & Time efficient automated responses with 24x7 availability.                       \\ \hline
Weakness      & Never guarantee a cost effective solution and always require expert supervision. \\ \hline
Opportunities & Strengthens the global economy with more promising and robust solutions.         \\ \hline
Threats       & May increase the unemployment ratio.                                             \\ \hline
    \end{tabular}}
\end{table}
\subsection{Social Control (contact less initiatives)}
Since this pandemic spreads via physical contact (even a simple touch) from the infected person or any surface, it is necessary to develop the habits of social control. With the exhaustive survey it was observed that various contactless initiatives are undertaken that involves work from home~\citep{NP110}, contactless - payment~|citep{NP111}, delivery~\citep{NP112}, drive-through COVID-19 testing~\citep{NP113}, etc. Social distancing solution is designed to combat the challenges that all essential workforces, and businesses eager to return to normal operations, are facing in the time of COVID-19. It is based on two basic concepts: passive and active. Passive approach is based on the interaction records of the workers (or any individuals) which utilizes the contact tracing in case an individual is tested positive, whereas in active approach, an alarm (via visual or audio signals) is generated for workers to adjust their current distance for following the social distancing protocol. Nowadays, there are many social distancing tools are available for automated monitoring. Landing AI~\citep{NP115.1} is a startup that provides companies with end-to-end AI platforms and has come up with a useful tool to monitor social distancing. Proximity Trace~\citep{NP115.2} is another solution that is based on device-to-device and device-to-gateway signaling without any dependency on client’s Wi-Fi or internet and location data (on site or off site). Spotr~\citep{NP115.3} also works as a safety and visibility tool for contact tracing under COVID-19 protocols. Furthermore, since we mostly contact the exposed parts of our body (eyes, ears, mouth, etc.) with our hands, many hands-free equipments are developed e.g. door openers~\citep{NP114}, biometric verification~\citep{NP116}, elevators~\citep{NP117}, smartphones, electrical and electronic household appliances (switch boards, fans, lights, bulbs, etc.), etc. These developments utilize the IoT enabled systems integrated with the AI technology to support the desired action. Most of these equipments make use of the combination of the mechanisms like voice-over-commands, action-on presence or motion and gesture-based user interaction~\citep{mewes2017touchless}. These contactless and hands-free schemes have been adopted in China, which has emerged as the only nation to be close to victory in the fight against COVID-19. With the example set by China, it is evident that these technological driven approaches and mechanisms can assist in the contactless or almost contactless (handsfree) initiatives to fight against COVID-19. Table~\ref{tab26} presents the SWOT analysis of above discussed procedures for social control during COVID-19 pandemic.

\begin{table}[]
    \centering
    \caption{SWOT analysis of the social control procedures in COVID-19.}
    \label{tab26}
    \resizebox{\linewidth}{!}{
    \begin{tabular}{|p{0.8in}|p{5.0in}|}     \hline        \textbf{SWOT} & \textbf{Remarks}\\         \hline
         Strength      & Energy efficient and hygienic environments are promoted to restrict the spread of infection. Helpful in curtailing the epidemic by reducing the chance of infection among high-risk populations.                                                       \\ \hline
Weakness      & Environment adaptability issues, need of self management, infrastructural overhead, requires expert supervision. Change in behavior of individuals, critical for elderly population.                                                                  \\ \hline
Opportunities & Encourages a viral free nation (even after post-COVID-19 world). Opening new avenues for smart-space infrastructure development. New standardization and patents are required. Healthier environment, automated surveillancing with limited manpower. \\ \hline
Threats       & Too much dependency on technology. Overhead is involved for any alternate adaptation. Privacy, security and trust issues.                                                                                                                             \\ \hline
    \end{tabular}}
\end{table}
\subsection{Tracking and tracing}
Contact tracing is the process to identify the people who came in contact with infectected person. Till date no vaccine is reported to cure COVID-19, hence to stop the spread of coronavirus, contact tracing is also one of the effective solutions. Table~\ref{tab27} indicates the different types of contact tracing methodologies. It is also evident that a minimum of 6 feet of social distance is sufficient enough to flatten the curve of COVID-19 spread~\citep{singh2020monitoring}.
\begin{table}[]
    \centering
    \caption{Different types of tracing technology.}
    \label{tab27}
    \resizebox{\linewidth}{!}{
    \begin{tabular}{|p{1.8in}|p{1.75in}|p{1.5in}|}
    \hline
         \multicolumn{1}{|c|}{\multirow{2}{*}{\textbf{Manual contact tracing}} }                                                                                                                                & \multicolumn{2}{|c|}{\textbf{Digital contact tracing} }                                                                                                                                                                                                             \\ \cline{2-3}
\multicolumn{1}{|c|}{}                                                                                                                                                                        & \multicolumn{1}{|c|}{\textbf{Location tracking}}                                                                                      & \multicolumn{1}{|c|}{\textbf{Proximity tracking} }                                                                                      \\  \hline
\begin{itemize}
  \item Healthcare workers interview an infected individual to learn about their movements and people with whom they have been in close contact.
  \item  Healthcare workers then reach out to the infected person’s potential contacts, and may offer them help, or ask them to self-isolate and get a test, treatment, or vaccination if available.
\end{itemize} & \begin{itemize}
\item Uses GPS data to determine individuals who were in the same place at the same time. \item Not accurate to indicate close physical contact (i.e. within 6 feet).
\item Accurate enough to expose sensitive, individually identifiable information about a person’s home, workplace, and routines.\end{itemize} & \begin{itemize} \item Uses bluetooth low energy (BLE) to determine whether two smartphones are close enough for their users to transmit the virus.\item Better than GPS or cell site location information.\item Closely identify proximity i.e. 6 feet.\end{itemize} \\ \hline
    \end{tabular}}
\end{table}
The European center for disease prevention and control (ECDC) found that time of exposure with a key threshold of greater than 15 minutes in a day for infection. All people who came in contact are treated as suspects and must be quarantined for at least 14 days because it takes 5 to 14 days for COVID-19 symptoms to appear. Every country is using their own mobile app for tracing the people effectively like: Arogya setu in India, Stopp Corona in Austria etc. Table~\ref{tab28} gives the details of some applications used by different countries along with the SWOT analysis shown in Table~\ref{tab29}.

\begin{table}[]
    \centering
    \caption{Details of Contract Tracing App.}
    \label{tab28}
 \resizebox{\linewidth}{!}{
    \begin{tabular}{|p{0.6in}|p{1.4in}|p{1.4in}|p{0.8in}|}
    \hline
        \textbf{Country}         & \textbf{Name}                                   & \textbf{Functionality}                        & \textbf{Platform}     \\ \hline
         
Angola          & Covid-19 AO                            & self diagnostic                      & Web          \\ \hline
Australia       & Coronavirus Australia                  & information, isolation registration  & Android, iOS \\ \hline
Austria         & Stopp Corona                           & contact tracing, medical reporting   & Android, iOS \\ \hline
China           & Alipay Health Code                   & contact tracing                      & Android, iOS \\ \hline
Czech Republic  & eRouška                                & contact tracing                      & Android, iOS \\ \hline
Finland         & Ketju                                  & contact tracing                      & Android, iOS \\ \hline
France          & ROBERT                                 & contact tracing                      & Unknown      \\ \hline
Germany         & {Ito}                              & contact tracing                      & Android      \\ \hline
Hong Kong       & Stay Home Safe                         & quarantine enforcement               & Unknown      \\ \hline
Iceland         & Rakning C-19                           & route tracking                       & Android, IOS \\ \hline
India           & Aarogya Setu                           & contact tracing                      & Android, iOS \\ \hline
Israel          & Hamagen                                & contact tracing                      & Android, iOS \\ \hline
Italy           & SM-COVID-19                            & contact tracing                      & Android, iOS \\ \hline
Netherlands     & PrivateTracer                          & contact tracing                      & Android, iOS \\ \hline
North Macedonia & StopKorona!                            & contact tracing                      & Android, iOS \\ \hline
Norway          & Smittestopp                            & contact tracing, route tracking      & Android, iOS \\ \hline
Poland          & ProteGO                                & contact tracing                      & Android, iOS \\ \hline
Russia          & Contact Tracer                         & digital contact tracing and alerting & Android      \\ \hline
Saudi Arabia    & Tawakkalna                             & curfew management                    & Android, iOS \\ \hline
Singapore       & TraceTogether                          & contact tracing                      & Android, iOS \\\hline
South Korea     & Self-Quarantine app                    & isolation registration               & Android, iOS \\ \hline
Sri Lanka       & COVID Shield                           & Self-Health Checking                 & Android      \\ \hline
United Kingdom  & NHS App                                & multipurpose                         & Android, iOS \\ \hline
United States   & coEpi                                  & self-reporting                       & Android, iOS \\ \hline
Vietnam        & NCOVI                                  & medical reporting                    & Android, iOS \\ \hline
Worldwide       & World Health Organization Covid-19 App & information                          & Android, iOS \\ \hline

    \end{tabular}}
\end{table}

\begin{table}[h!]
\centering
    \caption{SWOT analysis of tracking and tracing control strategies.}
    \label{tab29}
    \resizebox{\linewidth}{!}{
    \begin{tabular}{|p{0.8in}|p{5.0in}|}     \hline         \textbf{SWOT}& \textbf{Remarks}\\         \hline
         Strength      & Quick alert mechanism for users. Ensures more secure and robust COVID-19 prevention mechanism to end users.                                                     \\ \hline
Weakness      & Environment adaptability issues, involvement of the whole population is needed for more accurate results. Technological awareness is needed.                    \\ \hline
Opportunities & Technological advancements and platform compatibility could increase the effectiveness.                                                                         \\ \hline
Threats       & Dependency on technology hence sometimes adaptation is difficult. Malfunctioning of such applications may cause panic among users. Security threat is possible. \\
 \hline
    \end{tabular}}
\end{table}
\subsection{Technical expertise for maintenance or problem-solving}
Lockdown is the best way to stop the spread of coronavirus. People visiting sites and workplaces are more prone to infection, hence during lockdown engineers and technicians must avoid traveling and visiting factories, remote areas or crowded places~\citep{Jnt01}. However, it is not always possible to work remotely, specially for field jobs like coal mines, steel plants, construction sites etc. But IoT based smart equipment can reduce the involvement of people. Through smart sensors installed in machinery and augmented reality, remote monitoring of machines can be managed efficiently. With the help of smart sensors under the supervision of experts, it is possible to identify the health of machinery for future troubleshooting that helps to reduce the physical visits. SWOT analysis of the discussed method is detailed in table~\ref{tab30}.

\begin{table}[h!]
    \centering
    \caption{SWOT analysis of smart maintenance and problem-solving technique.}
    \label{tab30}
    \resizebox{\linewidth}{!}{
    \begin{tabular}{|p{0.8in}|p{4.0in}|}     \hline      \textbf{SWOT} & \textbf{Remarks}\\         \hline
         Strength      & Ensures efficient utilization of machineries and effective diagnosis of the system without affecting people.                     \\ \hline
Weakness      & Environment adaptability issues, technological awareness is needed.                                                              \\ \hline
Opportunities & Technological advancements and platform compatibility could increase the effectiveness.                                          \\ \hline
Threats       & Dependency on technology, hence sometimes adaptation is difficult. Security threat is possible. Unemployment ratio may increase. \\ \hline

    \end{tabular}}
\end{table}

\subsection{Dashboards}
The latest COVID-19 pandemic hits the world badly and everybody in this world is looking for the current status of the number of infections, deaths and other information associated with this infectious disease. In this regard diverse types of visualization have been provided to the users for specific requirements~\citep{sk64}. In epidemic thematic service map, real-time epidemic data can be visualized~\citep{mosenthal1990understanding} with the use of aggregation, heat and scattered points to display the epidemic trend in a manner of spatio-temporal evolution, and displays the development trend of the epidemic in multiple dimensions~\citep{bidhan2020mapping, sk66}. In pandemic situation awareness map visual representations of geographical distribution are generated based on the new confirmed cases, cumulative diagnoses and recovered cases maps. Epidemic correlation research and judgment map is used to find out the transportation information of COVID-19 diagnosis connected to the network through the real-time comparison of the coordinated traffic ticket information and epidemic observers. Mobile epidemic maps enable more people to view epidemic information conveniently and in real-time on portable devices.  Furthermore, spatial big data analytics can be applied for better insight and visualization along with the thematic applications of situational awareness of epidemic prevention and control. Related research findings and observations of epidemic situations can be provided to the concerned authorities and end users. Country specific authorities are displaying the data which helps not only for better planning but also in prevention and control of pandemic. Table~\ref{tab31} describes the SWOT analysis of the dashboard applications.
\begin{table}[h!]
    \centering
    \caption{SWOT analysis of dashboard applications.}
    \label{tab31}
    \resizebox{\linewidth}{!}{
    \begin{tabular}{|p{0.8in}|p{4.0in}|}     \hline      \textbf{SWOT} & \textbf{Remarks}\\         \hline
         Strength      & Effective way to see the current status of COVID-19. Ensures latest updates through multiple sources on a single platform. Easy to integrate with several heterogeneous sources. \\ \hline
Weakness      & Depends on the authorized sources, hence chances of duplicate information. Micro-level analysis is still missing in most of the applications.                                    \\ \hline
Opportunities & Technological advancements and platform compatibility could increase the effectiveness. Micro-level analysis could enhance transparency.                                         \\ \hline
Threats       & Dependency on technology and malfunctioning of such applications may cause severe panic. Presence of fake or false data is the biggest concern.                                  \\ \hline

    \end{tabular}}
\end{table}

\subsection{Supply chain management}
During the current pandemic, maintaining a healthy supply chain of essential goods is a challenging task. During this pandemic’s peak momentum, several delivery and logistic workers are risking their lives while delivering the  products door-to-door, because there is a high chance of exposure to viral infection~\citep{sk58}. Drones and robots (street bots) are more suitable replacements already implemented in some parts of the world~\citep{ye2020delivery}. The manpower could be used instead to operate and control these from close proximity, like local warehouses, without having to come into direct contact with people that may be infected. 

Supermarkets and other essential goods chains need to take extra measures to tackle the situation of panic during pandemic. A more promising way of supply chain management is needed to stop the blackmarketing of essential goods to enable smooth supply to the consumers. In supermarkets, smart shelves, smart fridges, video analytics, and an end-to-end connected supply chain can help retailers cope with uncertainty in their planning and even mitigate customers’ anxious behavior due to panic. Secure storage and transportation could play a vital role in the current situation to avoid black-marketing and hoarding of essential goods~\citep{chanyal2019black}. Table~\ref{tab32} presents the SWOT analysis of the measures of supply chain management measures during COVID-19.

\begin{table}[]
    \centering
    \caption{SWOT analysis of supply chain management.}
    \label{tab32}\
\resizebox{\linewidth}{!}{
    \begin{tabular}{|p{0.8in}|p{4.0in}|}     \hline      \textbf{SWOT} & \textbf{Remarks}\\         \hline
         Strength      & Effective supply-chain mechanism ensures availability of all essentials to people.                                                                 \\ \hline
Weakness      & Limited availability of transportation may lead delay in delivery to the customers                                                                 \\ \hline
Opportunities & Technological advancements are possible in terms of transportation and security.                                                                   \\ \hline
Threats       & Demand prediction of products is the most crucial issue as due to false prediction starvation, blackmarketing and hoarding will come into picture. \\ \hline
    \end{tabular}}
\end{table}

\subsection{Antitheft solutions for toilet rolls and antibacterial gels in public spaces}
Smart building solutions are the primary requirement of modern society~\citep{buckman2014smart}. Smart toilets and washrooms~\citep{plate2015smart} are very important to maintain hygiene. In a smart toilet, an IoT based toilet roll holder~\citep{kyser1989toilet} is an essential prerequisite for fair usage. Objective of smart toilet roll holders is to avoid unavailability of tissue papers and prevent the stealing of rolls. An IoT-enabled toilet paper holder that locks the paper and sends notifications when the paper is close to running out or notify if the holder is under threat could be a fantastic solution. Similarly, antibacterial gels dispensers are equally important to alert the cleaning staff for refilling and to avoid the stealing of gel. During COVID-19 pandemic, hygiene is the primary and most important concern to avoid the spread, therefore smart toilet rolls and antibacterial gel holders could play an important role to maintain the same. Table~\ref{tab33} highlights the SWOT analysis of the smart toilet and washrooms measures.

\begin{table}[]
    \centering
    \caption{SWOT analysis of smart toilet and washrooms amid COVID-19.}
    \label{tab33}
    \resizebox{\linewidth}{!}{
    \begin{tabular}{|p{0.8in}|p{4.0in}|}     \hline      \textbf{SWOT} & \textbf{Remarks}\\        \hline
         Strength      & Efficient way to maintain hygiene.                                                                                                                                                             \\ \hline
Weakness      & It is very difficult to deploy smart washrooms in rural and dense areas like at Dharavi, Mumbai (India). Washroom etiquette must be followed by all the people. Need more infrastructure cost. \\ \hline
Opportunities & Technological advancements are possible.                                                                                                                                                       \\ \hline
Threats       & False alarm may cause unnecessary panic. Maintenance is the biggest concern.                                                                                                                   \\ \hline

    \end{tabular}}
\end{table}

\subsection{3D Printing of medical supplies}
Due to abrupt acceleration of COVID-19 infections, diagnosis and treatment of patients become challenging because of limited health-care equipments. Chances of infection between corona warriors and patients during treatment depends on the use of personal protective equipment (PPE), gloves, face masks, air-purifier, goggles, face shields, respirators, and gowns. Due to the unexpected growth on demands of all these items, mass production is needed. But due to lockdown as a social distance measure, enough resources and manpower are not available, hence 3D printing came as a novel solution to produce these important items~\citep{livingston2020sourcing}. Our health systems were not designed to cope with a pandemic at this scale and, as a result, shortage of resources are now common across even the most advanced hospitals. 3D printing could be a lifesaver in the face of supply shortages caused by the coronavirus. 3D printers are capable of providing vital medical supplies quickly, such as ventilators, replacement valves. In Table~\ref{tab34} SWOT analysis is presented for the measures proposed with the use of 3D printing.
\begin{table}[]
    \centering
    \caption{SWOT analysis for the research works in 3D printing technology.}
    \label{tab34}
    \resizebox{\linewidth}{!}{
    \begin{tabular}{|p{0.8in}|p{4.0in}|}     \hline      \textbf{SWOT} & \textbf{Remarks}\\         \hline
         Strength      & User convenient, high product quality, low cost.                                                                                        \\ \hline
Weakness      & Costly process, sometimes long production time needed, material selection limitations, product quality depends on printer and material. \\ \hline
Opportunities & Customization of devices, technology and materials.                                                                                     \\ \hline
Threats       & Impact on environment, safety, threat to traditional workforce.                                                                         \\ \hline

    \end{tabular}}
\end{table}

\subsection{5G – for better connectivity}
Due to congestion and increased number of subscribers, present technologies (3G, 4G, etc.) and existing infrastructure are not sufficient enough for present and future requirements. During this deadly pandemic, an efficient and faster way of communication technology is desired for several personal, professional and social activities like online classes,  online professional meetings, online conferences, telemedicine, work-from-home, etc. Although still early in experimentation, 5G network is the latest development to provide better communication experience~\citep{ting2020digital}, which follows network slicing mechanisms to allocate the right networking resources in real-time and even in a predictive manner~\citep{alliance2016description}. In addition to above, a smart healthcare system is the need of today’s world, where an efficient way of communication system is required among all health workers, security people and administration, to manage this crisis without panic. A regional hospital, its ambulances and paramedics, and the city authorities can be interworked in one platform and receive updates in real-time. Therefore almost all personal and social activities are transforming to a virtual world with focus on online communication like conferences, training classes, presentations, meetings, etc. All these 5G based implementations are being tested in several countries like the UK, China, and the USA and it is expected to have it in commercial use in a few years. SWOT analysis  of 5G technology is discussed in Table~\ref{tab35}.

\begin{table}[]
\centering
    \caption{SWOT analysis for the research works in 5G technology.}
    \label{tab35}
    \resizebox{\linewidth}{!}{
    \begin{tabular}{|p{0.8in}|p{5.0in}|}     \hline      \textbf{SWOT} & \textbf{Remarks}\\         \hline
        Strength      & Effectiveness and efficiency, more bandwidth, high data rates, smoother hands-off,  Low latency. \\ \hline
Weakness      & Security issues, misuse of bandwidth, incompatibility with available hardware.                   \\ \hline
Opportunities & More intense security,  low energy consumption, high reliability.                                \\ \hline
Threats       & Security breach is possible, adaptability.                                                        \\ \hline
    \end{tabular}}
\end{table}

\subsection{Community services using drone}
COVID-19 has challenged and changed the personal and social life of people dramatically. During this unbreakable outbreak, contactless task accomplishment has become the most suitable solution to stop the spread of the virus. Drones are the latest disruptive force in the field of carriers. It is observed that drone technology is very useful in several application domains including healthcare~\citep{scott2017drone, kim2017drone, krey2019usage}. This section covers some use cases of drone applications like: announcements, monitoring and surveillance, drug delivery, etc.  against the COVID-19 outbreak.

In early April, 2020 at Manhattan, USA, a drone was monitoring the east river park area which was alerting people to follow the social distancing protocol~\citep{sk44}. It was named \enquote{the Anti-COVID-19 Volunteer Drone Task Force}, and suggested the crowd to maintain a minimum of six feet distance among each other. Similarly, in Kazakhstan, KazUAV drones are assisting the police department by patrolling the capital city~\citep{sk45}. These drones are equipped with the cameras and infrared sensors, monitoring people to follow lockdown measures. It is found effective  to identify multiple road bypass and irregular activities taking place in the target areas.

In India, IdeaForge, a leading manufacturer of UAVs has deployed drones to monitor public places in various regions of India~\citep{sk46} along with the thermal cameras to scan the body temperature of people over a large area~\citep{sk47}. Furthermore, GarudaUAV is helping local police forces to manage vehicular traffic and public overcrowding. While in Spain, France, and China drones are used to broadcast messages, requesting people to stay indoors, use masks before going out, etc. This helps to identify new likely cases without touching those who may be already infected. When it comes to deliveries, in China drones are deployed to ensure safe transportation of medical and other supplies in various places with minimal infection possibilities~\citep{sk48}. Along with it, drones are also used to deliver groceries and other essential items.

To enable sterilization, DJI~\citep{sk49} repurposed its agricultural drones to spray disinfectants at public places. Its’ Agras agriculture drone, which previously dispersed liquid pesticides, fertilizers, and herbicides on-farm plants, is now used to spray chlorine alcohol-based disinfectant in affected areas. Recently, Aertos 120-UVC, the first indoor drone with C-band ultraviolet (UVC) lights developed by Digital Aerolus to disinfect critical places like hospitals, market stores, airports and bus stops, etc. Table~\ref{tab36} presents the SWOT analysis of the measures of drone based approaches to fight COVID-19

\begin{table}[]
    \centering
    \caption{SWOT analysis for the research works in drone technology.}
    \label{tab36} 
    \resizebox{\linewidth}{!}{
    \begin{tabular}{|p{0.8in}|p{4.0in}|}     \hline      \textbf{SWOT} & \textbf{Remarks}\\         \hline
         Strength      & Easy deployment, reliable and robust.                                                                                                                                                 \\ \hline
Weakness      & Fuel and battery issues,  accidental collision, drones cannot carry heavier payloads and  deliver goods long distances.                                                                \\ \hline
Opportunities & Size, propulsion technology and coverage area could be improved, Surveillance in complex areas.                                                                                        \\  \hline
Threats       & Security breach is possible due to internet involvement. Hackers can hijack a drone using GPS jammers, drones may interfere with air traffic and cause confusion to commercial planes.  \\ \hline

    \end{tabular}}
\end{table}

\subsection{Innovative contributions developed in India}
Although India’s COVID-19 tally is increasing daily, researchers in India are taking no respite and coming up with ideas and innovations to fight the outbreak. Doctors, engineers, students and developers from startup firms are pitching-in to make the best use of technology to fight against COVID-19~\citep{SA201, SA202, SA203}. Table~\ref{tab37} presents some of the solutions developed in India to fight against the pandemic of COVID-19.

\begin{table}[]
    \centering
    \caption{Innovative solutions developed by various Indian researchers to fight COVID-19.}
    \label{tab37}
    \resizebox{\linewidth}{!}{
    \begin{tabular}{|p{1.1in}|p{4in}|}    \hline   
         Product                                & Features                                                                                                                                                                                                                                           \\
         \hline
Automatic mask creator machines        & Hygienic, fast, cost is expected to be 40\% cheaper.                                                                                                                                                                                               \\ \hline
Ruhdaar: The low-cost frugal innovator & A prototype of a low-cost ventilator, efficient and handy.                                                                                                                                                                                         \\ \hline
Jeeva Setu ventilators                 & An oven-sized, low-cost and portable ventilator.                                                                                                                                                                                                   \\ \hline
Low-cost PPEs: The Navy’s innovation   & A low-cost PPE with a special fabric with high ‘breathability’, which is suitable for hot and humid conditions prevalent in India.                                                                                                                 \\ \hline
Advanced wash basins                   & It allows the water tap and the soap dispenser to be mechanically operated without touch.                                                                                                                                                          \\ \hline
COVISAFE: Transporting patients        & To ensure the safe transport of COVID-19 patients, fits well on medical stretchers and is completely air-tight, equipped with oxygen supply and ventilators. When the coronavirus patient breathes, the air that comes out of the box is filtered. \\ \hline
Sanitizer Tunnels: A \enquote*{jugaad}          & India’s first organic disinfectant tunnel built in just two days using local materials and workforce, uses an organic fumigant instead of Sodium Hypochlorite, covered with polythene sheets.                                                      \\ \hline
Safe swab: Phone booth testing         & A phone booth coronavirus testing, fast and safe.                                                                                                                                                                                                  \\ \hline
UV sanitization device                 & Ultraviolet ray contactless sanitizers, both portable and stationary, cheaper.                                                                                                                                                                      \\ \hline
Robots                                 & To dispense hand sanitizer, deliver public health messages, to carry food and medicines in isolation wards.                                                                                                                                        \\ \hline
Microwave Sanitizer Atulya             & Fast, lightweight, portable, efficient and effective.                                                                                                                                                                                               \\ \hline

    \end{tabular}}
\end{table}

\section{Challenges, priorities and future directions}
\begin{figure}
    \centering
    \includegraphics[scale= 0.3] {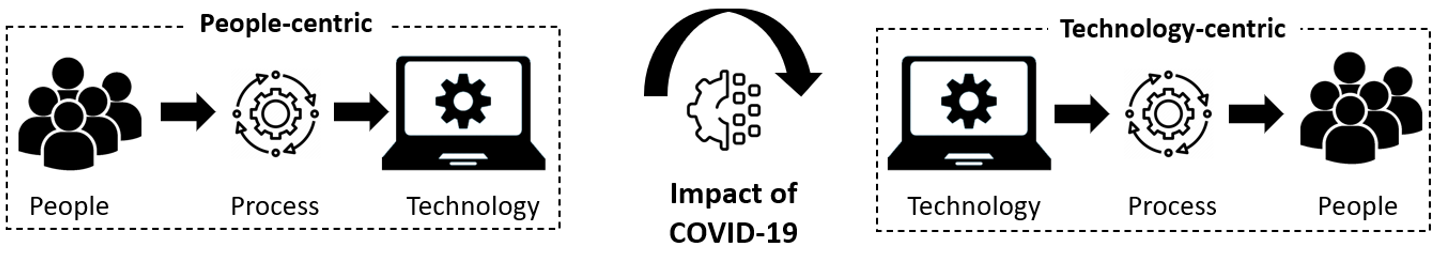}
    \caption{Impact of COVID-19 on the working environment.}
    \label{fig27}
\end{figure}

With social distancing related measures being the only possibility to fight this pandemic, it has resulted in diverse technological development across various domains displaying radical changes on the operating assets, and people; for instance, making the workspace, technological-intensive rather than people-intensive as shown in the Fig.~\ref{fig27}. Minimizing these impacts inculcate various challenges which needs to be tackled by setting certain priorities along with the future possibilities to achieve sustainable developments.

\subsection{Challenges}
The technological solutions are facing not only legal and regulatory challenges, but also socio-ethical dilemmas when applied in different sectors in the present context. A thorough study of these emerging technological solutions in the context of COVID-19 indicates that technology standalone cannot handle the crises but it can play a critical role in emergency responses, if proper policies and regulatory measures have been fixed up. We here analyse the challenges from four main aspects: regulatory consideration, people’s privacy preservation, security concerns, and the lack of unified databases.

\subsubsection{Regulatory consideration}
The use of emerging technologies in the healthcare sector to fight against coronavirus should be considered carefully with regulatory laws. While the features of various technologies can bring benefits, it also pose a legal and regulatory challenge if there is no party that is responsible and can be held accountable. For example, in the blockchain network, it will be important to consider what law might apply to transactions and what appropriate risk management should be put into place. Likewise, in the case of AI, it might be easier to create standard norms of legal schemes and internal governance models that will dictate the governing law for AI operations in healthcare. Specially, we should also consider legal issues about sharing content and personal information, with copyright, infringement and defamation.

\subsubsection{People’s privacy preservation}
In the coronavirus tracking applications, privacy protection of people is highly important. The governments can use mobile location data to track the outbreak spread, but this solution must ensure the privacy of user data, especially sensitive information, such as home address, banking details, shopping records, etc. The government agencies may impose privacy laws on the user tracking mobile apps to ensure the safety and privacy of the public. Nowadays many healthcare organizations and institutions are collecting data from their patients via electronic healthcare records that helps monitor the COVID-19 disease symptoms and respective treatments. In such healthcare activities, the conflict between data collection and user privacy is inevitable that needs to be solved by laws and reinforcement from the related authorities.

\subsubsection{Security concerns}
COVID-19 tracking and monitoring applications require secure monitoring platform such as blockchain, to ensure safety and privacy of the individuals. However, many stakeholders concern inherent security weaknesses in blockchain, especially related to financial and healthcare issues. It is also one of the concern that data threats or adversaries can enter the blockchain to hold its control, which can lead to serious consequences, like modifications of medical transaction, patient records, thereby further raising the privacy concerns. Security is also a sensitive issue of AI systems for healthcare applications, where malicious attacks can inject false data or adversarial sample inputs and subsequently affect the performance of the system. Therefore, security issues should be given high priority during the design and development of any technological solution during this pandemic.

\subsubsection{Lack of unified databases}
A critical challenge in the coronavirus fighting is the lack of unified database related to epidemic such as infected cases, affected areas, and medical supply status. Most of the current coronavirus related databases come from several resources, i.e. social media, health care organizations, etc.; but it is not sufficient for data driven applications, desired to create greater impacts on COVID-19 fight. Countries are sometimes reluctant to share data, therefore it is challenging to understand the epidemic characteristics for proactive measures. It is difficult to share appropriate public health guidelines without local and global data availability.

\subsection{Priorities}
This pandemic situation has shattered every dimension that defines our world. In order to sustain our living with minimal loss following are the certain set of priorities defined by government agencies and policymakers which need to be followed:
\subsubsection{Flatten the curve}
\begin{figure}
    \centering
    \includegraphics[scale= 0.35] {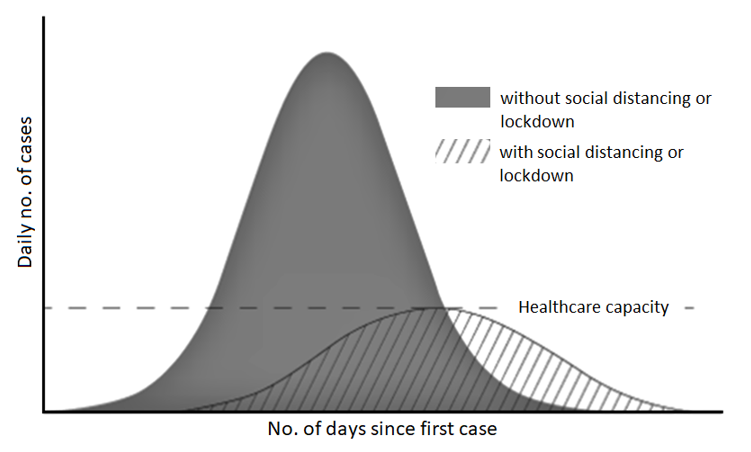}
    \caption{Social distancing leading to flatten the curve.}
    \label{fig28}
\end{figure}

This means to reduce the number of critical care patients to sustain within the healthcare capacity as shown in Fig.~\ref{fig28}. With no availability of the concrete solution aggressive social distancing (lockdown), quarantine, and self-hygiene are the only possible remedies. Technological advancements have been contributing continuously to develop automated approaches that can aid in monitoring such measures.

\subsubsection{Survivability and resilience}
To sustain in the post COVID-19 world, it is necessary to ensure the survivability and recoverability of every value link in the chain of development that includes organizations, agriculture, society, etc. This is possible if all firms unite to contribute in meeting the necessary demand to fight against this pandemic, while improving productivity and keeping the long-term relationship among each other instead of deliberately focusing on the economy.

\subsubsection{Digital advancements}
Moving towards digital business models with the aim to meet the growing needs, will encourage to set up the workspace with AI technology fused with human ingenuity. These advancements must focus on value creation, rather than being predominantly technology driven (with human supervision). This would lead towards a contactless working environment while also improving the productivity and efficiency of the firms.

\subsubsection{Handling ecological and environmental threats}
Looking back to the past, the last century tolerated extreme upheavals in the form of five dangerous virus attacks that are far from rare. Coronavirus is one of the most impactful events also causing climate change, over stressed population and rising worldwide inequality. Here it would be admirable if technical initiative could handle these unimaginable events. Following the recent trends, disruptive technologies complemented with a number of computer-assisted algorithms can prevail to minimize the negative impact of COVID-19. As a use case, stakeholders may also utilize AI and machine-learning methods to enumerate political, economical, ecological and environmental risks grounded on social-media/image/hyperspectral data analytics in the progression of strategic decision making.

\subsection{Future directions}
\begin{figure}
    \centering
    \includegraphics[scale= 0.45] {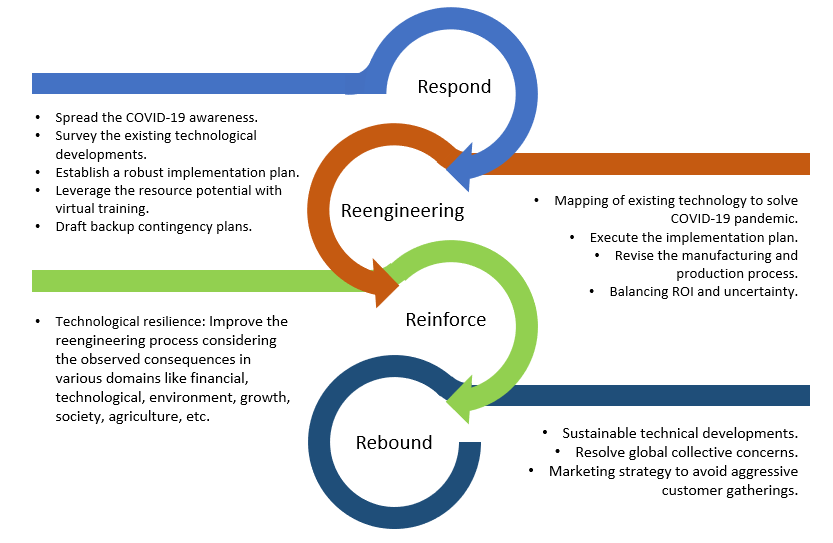}
    \caption{4Rs of technological action plan to fight COVID-19.}
    \label{fig29}
\end{figure}

2020 has brought an unanticipated set of challenges for economies around the world, more likely there will be a climax of an impending slump stepping up throughout 2019. Apparently seems to be an unending COVID-19 pandemic - individuals, organizations, nations, the whole world is affected. Thus, there is an urgency to accelerate responsible action plans to overcome this crisis and also turn it into something positive for the society at large and for generations to come. COVID-19 pandemic situation demands a serious contribution from policymakers and regulators to demonstrate legal plausibility, ethical soundness, and effectiveness of the deployment of emerging, future and disruptive technologies under time pressure. The action plan can rely upon 4 horizons: respond, reengineering, reinforce, and rebound as highlighted in Fig.~\ref{fig29}.

In the first stage, it is highly imperative to respond proactively and act decisively in order to have the best possible use of existing and emerging technologies to fight against the COVID-19 pandemic. It will help to maintain continuity, productivity and mental well-being of the individuals so that they can contribute efficiently as per the need. In the case of COVID-19, this stage is either undergoing or already over in most of the countries. Once the immediate emergencies are covered up through the respond stage, the next is to seek a serious effort to review and reorganise the shape of technology-based solutions under the re-engineering stage. It covers established strategies, detailed plans and structural reformation of rules and guidelines to re-frame the existing setup designed to contain the spread of COVID-19. It is also a time to realign priorities and allocate resources for each and every technology sector along with balancing between cost and benefits. The third stage is the reinforce stage which includes sufficient time to strengthen and build technological capabilities on the basis of previous outcomes. At this stage, to meet the COVID-19 challenges, the convergence of technology could be performed with efficient process workflows and automation. The last stage would be rebound, a time to drive rapid growth and sustainable development to gain the momentum in all sectors either conservatively, moderately or aggressively to steer it in the right direction for the future. Following are some of the technology-enabled future research directions for fighting COVID-19 pandemic:

\subsubsection{Embracing touchless technology}
This pandemic has set a benchmark to lead this world towards touchless technological advancements via AI, IoT, data analytics, machine learning, etc. The advancements are focused on analysing the commands set by the human gesture, motion, voice, expressions, retina, holographic imaging, etc., where each activity can be mapped with a certain set of actions. For instance, smart voice-enabled assistants led by iPhone (Siri), Amazon (Alexa), Google assistant, interpret the speech commands to perform the desired action like make a call, search, set an alarm, take a picture, etc., facial feature recognition for person identification via smart cameras and sensor devices, holographic projections can aid in the touchless stimulation of the same set of commands as by the touch-enabled technologies.

\subsubsection{Optimize blockchain for better deployment}
Blockchains are computationally expensive and involve high bandwidth overhead and uneven delays. To fight COVID-19 pandemic, a smarter healthcare system is required for patient monitoring, telemedicine, drug supply, etc., in which an optimized blockchain platform is necessary for reduced network latency, increased throughput, and improved security. More specifically, lightweight blockchain designs in healthcare are necessary to optimize data verification and transactional communication for ultra-low latency information broadcasting. Seeking for an optimized blockchain solution, the size of blockchain can be minimized by establishing local and private blockchain networks to monitor the outbreak in a certain area for the fast response.

\subsubsection{Accelerate AI developments}
During this COVID-19 pandemic, smart healthcare is the primary need that involves telemedicine, robotic services, medicine delivery, diagnosis, etc. In this present era of artificial intelligence, healthcare data can be better understood via advanced machine learning algorithms. Conventional AI approaches are capable of providing smart solutions to existing disease up to some extent but seem inefficient for pandemic situations, hence more robust algorithms and techniques are required for the present combat against COVID-19. There is a need to develop adaptive AI architectures, specialized in medical applications that could empower future intelligent data analytics with the ability to handle multi-source healthcare data.

\subsubsection{Technological convergence for robust development}
Security and privacy are the primary concerns of every existing technique. To fight against COVID-19, secure technological convergence is required for safe and effective utilization of recent innovations in a public environment. For instance, AI along with blockchain-enabled security can be incorporated with other technologies like IoT, data analytics, etc. to build a more robust and effective system. Availability of massive storage capacity and high computation capability along with blockchain based security could play an effective and efficient way to process data related to sample collection, telemedicine, and drug delivery. Meanwhile, the tracking data of infected people is also to be kept confidential and secure enough from any breach to avoid panic during such pandemic. Hence, development of a more secure and transparent tracking system is required in future. Likewise, the transportation system must also be secure enough to fulfil the people's trust and confidence.

\subsubsection{Strengthen digital infrastructure}
The worldwide COVID-19 pandemic demands development of new capabilities and more cultural power to embrace the transition. Now there is a need to establish optimized technology and environment plans for empowering remote working culture and bridging the communication gap among all stakeholders. It is also crucial to evaluate networking capabilities, accelerate device deployment, and leverage virtual environments for a sustainable future. Developing a distributed work environment upon strong technological backbone and leveraging a secure network can help swiftly navigate the crisis with minimum loss.

\subsubsection{Real-time monitoring of IoT and big data applications}
One of the advantages we have today in the battle against COVID-19 is the availability of massive storage, extreme computing capability and big data processing tools. COVID-19 case details are continuously floated by all countries to perform real-time data analytics for improved, actionable insights with help of robust big data tools which would aid in preparing the necessary action plan to secure current and future needs. Technological advancements like contactless biometric and IoT enabled surveillance systems have been playing an effective role during this pandemic to establish post COVID-19 requirements. Big Data tools could also be supportive in studying the metrics about population movements, public compliance, health monitoring, etc. to attain the objective of social distancing and lockdown.

\section{Conclusion}
In this article, a state-of-the-art survey on the utilization of emerging, future and disruptive technologies to combat the coronavirus (COVID-19) pandemic is presented. We have first introduced the concept of disruptive technologies and its scope in terms of human-centric and smart-space approaches. We classified various emerging and future technologies either as human-centric or smart-space categories. In the battle against this outbreak, the present article highlights actively leveraged detailed list of digital technologies such as artificial intelligence (AI), big data, cloud computing, blockchain, 5G, etc., which have exceptionally improved the efficiency of the country’s efforts in epidemic monitoring, tracking, prevention, control and treatment, and resource allocation along with the SWOT analysis of the discussed techniques. Hence, this article acts as the foundation for various stakeholders to pursue and convert opportunities into strengths and prevent weaknesses from turning into threats.

Later, it also covers the associated challenges, insight priorities, and effective future directions. The current COVID-19 pandemic challenges the whole world in all dimensions and still, there is no assurance of permanent process to stop the spread, even it is also not possible to predict the day of independence from this deadly coronavirus. People’s privacy and data security along with regulatory consideration are the most challenging issues during this combat with COVID-19. Meanwhile, priorities during the pandemic have been discussed to attract the global community for the betterment of mankind. At last, the future directions have been discussed which offers the most suitable and workable areas to researchers, scientists, and health workers. It is believed that these future directions will transform the existing infrastructure very soon and the world will become capable enough to fight this type of crisis in the future.

The health sector is the frontline accountable community to fight against any pandemic, but the discussed technologies definitely enhance their capability and efficiency. It is obvious that technology cannot prevent nature originated pandemics; but can prevent the spread of the virus, and educate, alert, and empower the front-row warriors and society to be aware of the situation to reduce the impact for the well-being of humanity. In the end of this article, we are sure that enough data is getting generated around the world about this pandemic, in the form of huge numbers of  patients’ profiles, drug test reports, legal frameworks, social churning analysis reports, financial costs, reports from medical, social, governance, financial and technological experts, that the use of the above discussed emerging technologies would be feasible for the purpose which will help the scientists build the digital ecosystem to track the virus eruption, insulate the population from getting infected and help the people coexist with the epidemic in the shortest possible time.

\bibliographystyle{apa}\biboptions{authoryear}
\bibliography{reference}

\end{document}